# An Adaptive Physics-Driven Deep Learning Framework for a Two-Phase Stefan Problem

Meraj Hassanzadeh[1], Ehsan Ghaderi[1], Fatemeh Fatahi[1], and Mohamad Ali Bijarchi[*1]

[1]Department of Mechanical Engineering, Sharif University of Technology, Tehran, Iran

## Abstract

Thermal Energy Storage (TES) using Phase Change Materials (PCMs) represents a critical technology for sustainable energy management and grid stability. This study presents a novel Physics-Driven Deep Learning (PDDL) framework for modeling the complex solid-liquid phase transition in a two-dimensional PCM-based TES system integrated with finned heat exchangers. The system operates under transient forced convection with cooling air, presenting a challenging Moving Boundary Problem (MBP) characterized by intricate phase interface dynamics and strong geometrical dependencies. Conventional numerical methods for such Stefan problems face significant computational burdens due to repeated meshing requirements at the evolving interface. To overcome these limitations, we develop a multi-network PDDL approach that simultaneously predicts the solid phase temperature field, fin temperature distribution, and the moving phase boundary position. The architecture employs three specialized deep neural networks operating in parallel, constrained by the governing physical laws of energy conservation and interface conditions. Comprehensive validation against established analytical benchmarks demonstrates the framework's exceptional accuracy in predicting interface evolution and temperature distributions across various aspect ratios. The model successfully captures the parametric influence of geometrical parameter on solidification rates and thermal performance without requiring mesh regeneration. Our approach provides an efficient computational paradigm for optimizing PCM-based TES systems and offers extensibility to three-dimensional configurations and multi-material composites, presenting significant potential for advanced thermal energy system design.

---

[*] corresponding author: bijarchi@sharif.edu

## 1. Introduction

Thermal energy storage (TES) has garnered significant attention as a critical technology for enhancing energy efficiency and enabling the integration of renewable energy sources. Energy storage through latent heat storage during the melting and solidification processes of Phase Change Materials (PCMs) occurs in various fields, including solar water heating [1], electronic equipment cooling [2], and thermal management for battery [3]. One of the notable advantages of PCMs is the amount of energy they store or release during the phase change process. The high storage capacity during an isothermal process has made latent heat storage preferable to sensible heat storage [1]. In PCMs, the phenomenon of heat transfer is nonlinear and transient, and the heat transfer problem accompanied by phase change involves a moving solid–liquid interface, which is known as the Moving Boundary Problem (MBP) [4-5]. The challenge lies in the nonlinear nature of the MBP, which limits analytical solutions to only simple geometries and boundary conditions [6-8]. One of the analytical solutions for the one-dimensional MBP, proposed in the nineteenth century and developed thereafter, is known as the Stefan problem. Stefan formulated and presented the motion of the solid–liquid interface and the temperature distribution in a layer of freezing water [9].

In recent years, the scientific literature on modeling heat transfer during melting and solidification processes has been developed [10-12]. Zalba et al. [13] conducted a comprehensive review of thermal energy storage involving the MBP and presented models and numerical solutions. Zivkovic and Fujii [14] simulated the transient behavior of isothermal phase change with simple computations and proposed a mathematical model based on enthalpy, which showed good agreement with experimental results. Esen [15] developed a two-dimensional simulation model to describe the transient behavior of a phase change unit using the enthalpy-based finite difference method, and experimentally as well as theoretically investigated a cylindrical phase change storage tank connected to a solar heat pump system for space heating. Another challenge in using PCMs for thermal energy storage is their relatively low thermal conductivity, and numerous studies have been conducted so far to enhance the heat transfer rate in latent heat storage systems. Lamberg and Siren [16] used internal fins to improve heat transfer in a confined PCM storage system and presented a simplified analytical model that investigates the position of the solid–liquid interface and the temperature distribution in the fin during the solidification process, with a constant end-wall temperature in a two-dimensional finned PCM storage system. Talati et al. [17] investigated horizontal plate-type internal fins in a rectangular PCM storage unit to improve heat transfer and compared this analytical model with a two-dimensional numerical solution based on the enthalpy method for validation. As a result, considering the aforementioned studies, PCM phase change in latent heat storage systems has been extensively investigated experimentally, analytically, and numerically [18-21].

Machine Learning (ML) and Artificial Neural Networks (ANNs) have long attracted the attention of researchers in a wide range of scientific and engineering applications [22-23]. Recently, ANNs were used to evaluate convective heat transfer coefficients by training on experimental data obtained from liquid crystal thermography and transient conduction analysis [24]. In another study Kim and Lee [25], Convolutional Neural Networks (CNNs) were employed to predict turbulent heat transfer based on wall information obtained from Direct Numerical Simulations (DNS). Therefore, over the years, given the availability of large volumes of data, researchers have developed data-driven models due to their superior performance; however, these models required historical data and lacked the capability to learn beyond the dataset [26]. In recent years, the application of Physics-Informed Neural Networks (PINNs) for solving various and complex problems in different scientific and engineering fields, particularly in fluid mechanics and heat transfer, has gained considerable attention [27-28]. In 2019, Raissi et al. [29] introduced PINNs as an innovative and new framework in deep learning. Unlike traditional machine learning methods, which are data-driven, PINNs rely mainly on physical constraints, incorporating governing physical equations and boundary conditions directly into the neural network's loss function. This approach reduces dependence on large datasets, whose collection is costly and difficult [30-31]. Moreover, PINNs do not require time-consuming meshing processes and are mesh-free.

The efficiency of PINNs in parametric conditions is notable, as they can approximate responses for different input values within a single training process, unlike traditional Computational Fluid Dynamics (CFD) methods, which require numerous costly simulations, thus making PINNs an efficient and powerful tool for optimization [32]. Computationally, the parametric surrogate model can reduce computation time by about 14% compared to traditional CFD methods [33]. For example, Lu et al. [34] used an innovative inverse design approach for finned heat sink systems and demonstrated the application of PINNs in conjugate heat transfer and fluid mechanics. They employed a specialized hybrid PINN to parameterize geometric and operational inputs, enabling the identification of optimal heat sink designs by backtracking from desired targets. Ghaderi et al. [35] specifically investigated the ability of PINNs to learn the magnetic field response as a function of design parameters in two-dimensional (2D) Magneto Hydro Dynamic (MHD) problems and proposed an innovative framework for efficient analysis and optimization. The physical and geometric parameters are injected into the network input, and the capability of the PINN method in addressing parametric and inverse problems is demonstrated. A parametric solution for problems related to PCM has not yet been explored in the existing literature. However, research has been conducted on the application of PINNs in PCM-related studies [36-38].

As the literature review indicates, the use of fins to enhance heat transfer in TES systems has been a subject of interest. However, no previous study has addressed this problem using PINN methods. The moving boundary inherent in this problem's physics presents a challenge, as standard PINN solvers are inadequate, necessitating modifications to the network architecture. Furthermore, the time-dependent nature of the physics and the known challenges of PINNs in such problems require methodological adaptations. In this study, we aim to simulate the solid-fluid moving boundary problem in TES by modifying the network structure and employing adaptive techniques, such as adaptive sampling and adaptive weighting. The geometric parameter, defined as the ratio of fin length to cell height (where a cell comprises half of the fin and half of the PCM), is incorporated as a variable input into the learning process, and simulations are conducted for various values of this parameter. The proposed method is designed to predict the solid-liquid interface position, fin temperature, and temporal evolution of the solid fraction for different values of this geometric parameter.

## 2. Problem Description

The application of solidification in PCM was investigated in the studies by Mosaffa et al. [18] and Kothari et al. [39], with the corresponding geometry illustrated in Figure 1. Given that the liquid was initially at the solidification temperature, only the solid temperature and fin temperature within the PCM were considered as functions of time and space, and the liquid temperature was assumed to be constant. The problem was modeled in two dimensions and heat conduction mode was considered in this study, while the material properties were assumed to be constant. The geometric dimensions of the problem are provided in Table 1. Due to symmetry, only one-quarter of the domain was analyzed to reduce computational costs. Paraffin is utilized as PCM while the fin is made from Aluminum and the relevant physical properties for the problem under investigation are provided in Table 2.

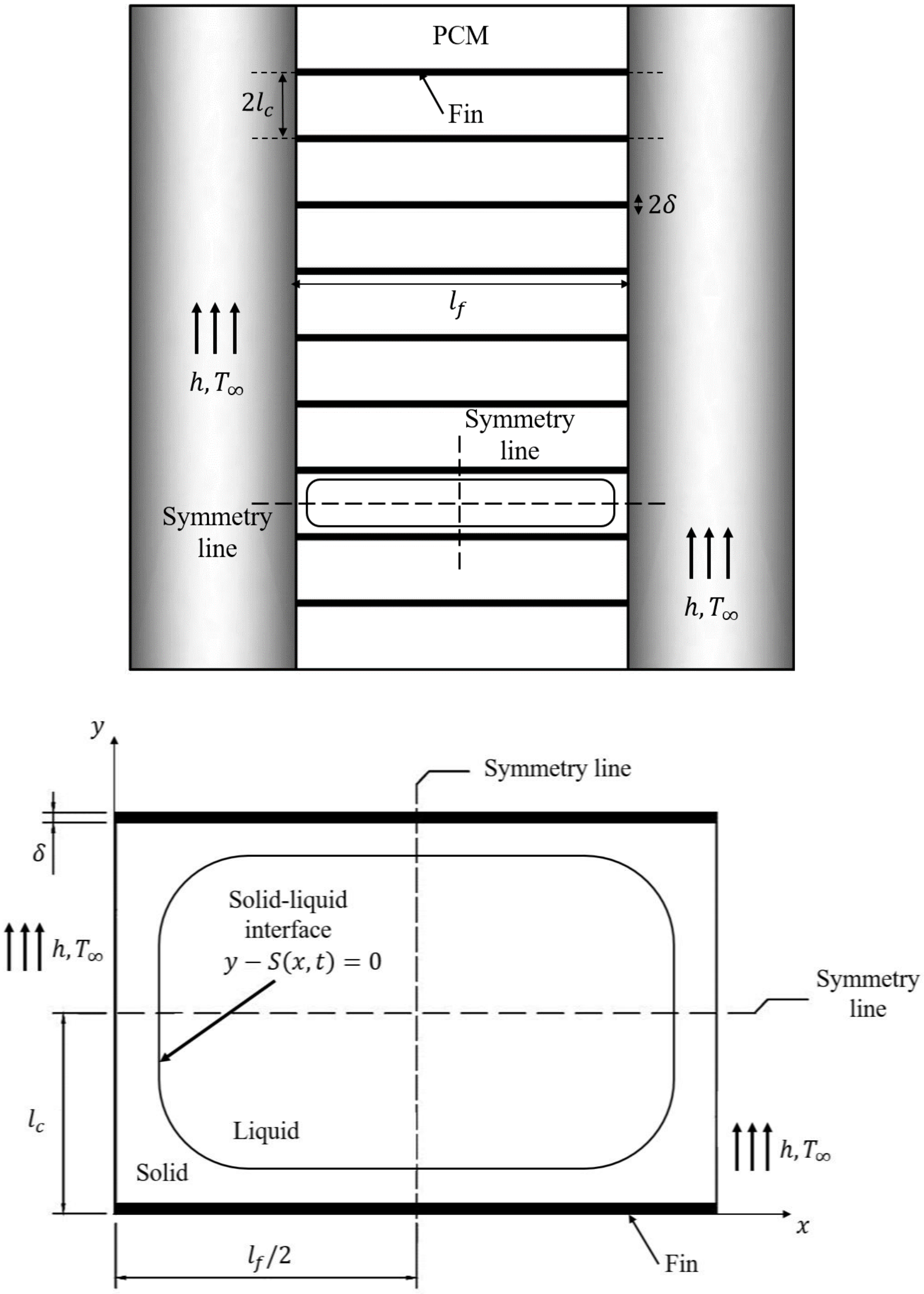

**Figure 1.** Schematic of the problem domain

**Table 1.** Geometrical specifications of the PCM and the fin.

| **Notation** | **Definition** | **Value** |
|---|---|---|
| $l_f$ | Length of fin | [5-25] mm |
| $l_c$ | Half height of cell | 10 mm |
| $\delta$ | Half thickness of fin | 0.5 mm |
| $\mathrm{P} = \frac{l_f}{l_c}$ | Cell aspect ratio | $\mathrm{P}_{min} = 1, \mathrm{P}_{max} = 5$ |

**Table 2.** Thermo-physical properties of the PCM and fin.

| **Property** | **Paraffin (PCM)** | **Aluminum Fin** |
|---|---|---|
| Density, $\rho$ (kg/m$^3$) | 830 | 2770 |
| Specific heat, $c$ (J/kg-K) | 1920 (solid)<br>3260 (liquid) | 875 |
| Thermal conductivity, $k$ (W/m-K) | 0.514 (solid)<br>0.224 (liquid) | 177 |
| Latent heat of fusion, $L$ (J/kg) | 251000 | - |
| Melting/solidification temperature, $T_m$ (°C) | 32 | - |

Given the geometry illustrated in Figure 1 and the stated assumptions, the governing equations for the PCM and the fin simplify to Equations (1) and (2):

$$\frac{\partial^2 T_s}{\partial x^2} + \frac{\partial^2 T_s}{\partial y^2} = \left(\frac{\rho c}{k}\right)_s \frac{\partial T_s}{\partial t} \quad (1)$$

$$\frac{\partial^2 T_f}{\partial x^2} + \frac{\partial^2 T_f}{\partial y^2} = \left(\frac{\rho c}{k}\right)_f \frac{\partial T_f}{\partial t} \quad (2)$$

In these equations, $T_s$ and $T_f$ represent the temperatures of the solid PCM and the fin, respectively. Known the nature of the problem under investigation, both of these temperature fields are functions of the spatial coordinates $(x,y)$ and time $(t)$. Given the physics of the problem and its governing equations, appropriate initial and boundary conditions are required for the simulation. The boundary conditions governing the problem are given by Equations (3) to (11). The specific formulation of Equations (6) to (9) is a direct consequence of the geometrical configuration adopted, as illustrated in Figure 2(b), which exploits the inherent symmetry of the problem to simplify the model. Furthermore, in these equations, $h$ represents the convective heat transfer coefficient with an assigned value of 65 W/m²-K, and $T_\infty$ denotes the ambient air temperature, set at 10°C. Since the problem is classified as a MBP, the interface between the phase-change material phases evolves with time. This dynamic introduces an additional unknown into the system, consequently necessitating an extra boundary condition for moving interface to achieve a well-posed formulation. This boundary condition is expressed in Equations (10) and (11) to achieve a closed system of equations.

$$k_s \frac{\partial T_s(0,y,t)}{\partial x} = h(T_s(0,y,t) - T_\infty) \tag{3}$$

$$k_f \frac{\partial T_f(0,y,t)}{\partial x} = h\left(T_f(0,y,t) - T_\infty\right) \tag{4}$$

$$T_s(x,\delta,t) = T_f(x,\delta,t) \tag{5}$$

$$\frac{\partial T_s}{\partial x}\left(\frac{l_f}{2}, y, t\right) = 0 \tag{6}$$

$$\frac{\partial T_s}{\partial y}(x, l_c, t) = 0 \tag{7}$$

$$\frac{\partial T_f}{\partial x}\left(\frac{l_f}{2}, y, t\right) = 0 \tag{8}$$

$$\frac{\partial T_f}{\partial y}(x, 0, t) = 0 \tag{9}$$

$$T_s(x, S(x,t), t) = T_m \tag{10}$$

$$\left(1 + \left(\frac{\partial S}{\partial t}\right)^2\right)\left(k_s\left(\frac{\partial T_s}{\partial y}\right)\right) = \rho_s L \frac{\partial S}{\partial t} \tag{11}$$

The initial conditions governing the PCM, the fin, and the moving interface ($S$) are specified in Equations (12) to (14).

$$T_s(x,y,0) = T_m \tag{12}$$

$$T_f(x, y, 0) = T_m \tag{13}$$

$$S(x, 0) = 0 \tag{14}$$

In addition to the introduced equations, a one-dimensional model has also been considered in this study for comparative purposes. For a detailed examination and a more comprehensive understanding of the underlying physics, please refer to Supplementary.

## 3. Solution Methodology

In this section, the chosen methodology for this research as illustrated in Figure 2, is explained. This approach leverages the governing physical equations, along with the initial and boundary conditions, to facilitate learning without relying on any pre-existing CFD data. Unlike conventional numerical methods, this technique eliminates the need for equation discretization or mesh generation within the solution domain. Instead, points are selectively sampled within the physical domain, and the neural network optimization process is conducted directly at these locations. Given the physics of the problem under investigation, it was essential to generate these points adaptively to accurately capture the moving boundary. Consequently, as depicted in Figure 2, the input layer of the neural network incorporates adaptively selected and dynamically varying spatial coordinates ($x^*$,$y^*$), temporal coordinate ($t^*$), and geometric parameter ($\mathrm{P}^*$). A key point to emphasize is that the geometric parameter in this study was treated as a variable input to the network, enabling its integration into the learning process. Thus, upon a single training execution, the proposed method is capable of delivering predictions for various values of the $\mathrm{P}^*$ parameter. Due to the coupled nature of the governing equations, three parallel neural networks were employed to simultaneously predict the dimensionless solid temperature and heat fluxes ($T_s^*, q_{s,x}^*, q_{s,y}^*$), fin temperature and heat fluxes ($T_f^*, q_{f,x}^*, q_{f,y}^*$), and the moving interface position ($S^*$). A single network is insufficient for this system because the interface depends only on $x^*$ and $t^*$, whereas the temperature fields depend on the full spatial domain ($x^*$,$y^*$) and $t^*$. Additionally, the fin is significantly thinner than the PCM region; therefore, using a single shared network for both domains could cause the learning process to underrepresent the fin compared to the PCM, compromising accuracy.

### 3.1. Pre-training Strategy

A pre-training strategy was implemented using one-dimensional analytical solutions from [18] to initialize the networks. This approach provides the network with physical intuition regarding interface dynamics and temperature distributions, offering significant advantages over conventional initialization methods like Xavier initialization. The pre-training ensures the network begins with physically meaningful approximations, accelerating convergence and improving solution accuracy.

### 3.2. Adaptive Sampling Algorithm

An enhanced adaptive sampling technique dynamically updates collocation points during training based on the evolving interface position. This method serves two critical purposes:

1. Interface Resolution Enhancement: By concentrating sampling density near the moving interface, the algorithm captures sharp gradients and complex interface dynamics with improved accuracy.
2. Domain-Aware Point Allocation: The method automatically distinguishes between solid and liquid regions, assigning collocation points to the appropriate phase-specific governing equations, ensuring physical consistency throughout the optimization process.

The sampling strategy incorporates curvature-based weighting, prioritizing regions with high interface complexity to maximize computational efficiency.

### 3.3. Adaptive Loss Weighting

An adaptive weighting mechanism dynamically balances the contribution of different physical constraints during training. This approach addresses several challenges inherent in multi-physics PDDL optimization:

Loss Scale Disparity: Different physical terms naturally exhibit varying magnitudes, potentially dominating the optimization process

Training Dynamics: The relative importance of boundary conditions, interface conditions, and PDE residuals evolves during training

Convergence Stability: Fixed weights often lead to imbalanced constraint satisfaction and suboptimal solutions

The weighting scheme employs an exponential formulation with moving average updates the equation $w_k^t = \tau w_k^{t-1} + (1-\tau)\frac{\mathcal{L}_k^t}{\mathcal{L}_k^{t-1}}$, where $w_k^t$ represents the weight for loss component k at timestep $t$, $\tau = 0.8$ controls the update smoothing, and $\mathcal{L}_k^t$ denotes the individual loss terms at timestep $t$. This formulation automatically increases weights for struggling loss components while reducing emphasis on well-satisfied constraints.

### 3.4. Hybrid Optimization Strategy

A two-phase optimization approach combines the complementary strengths of Adam and L-BFGS algorithms:

Adam Optimization Phase: The Adam optimizer conducts initial training for 20,000 epochs, efficiently navigating the high-dimensional parameter space to establish approximate solutions for both temperature fields and interface position. This phase benefits from Adam's robustness to noisy gradients and adaptive learning rates.

Adaptive L-BFGS Refinement Phase: Following initial convergence, the algorithm transitions to L-BFGS optimization with a cyclic adaptive sampling strategy. Each L-BFGS cycle begins with point cloud updates based on the current interface prediction, followed by multiple L-BFGS iterations on fixed collocation points. This hybrid approach maintains optimization stability while accommodating interface evolution.

Algorithm 1 summarizes the complete training workflow, which integrates pre-training, adaptive sampling, loss weighting, and hybrid optimization to achieve robust convergence for the challenging two-phase Stefan problem with moving interface. The implementation details of the proposed methodology, corresponding to this research, are available in the GitHub repository.

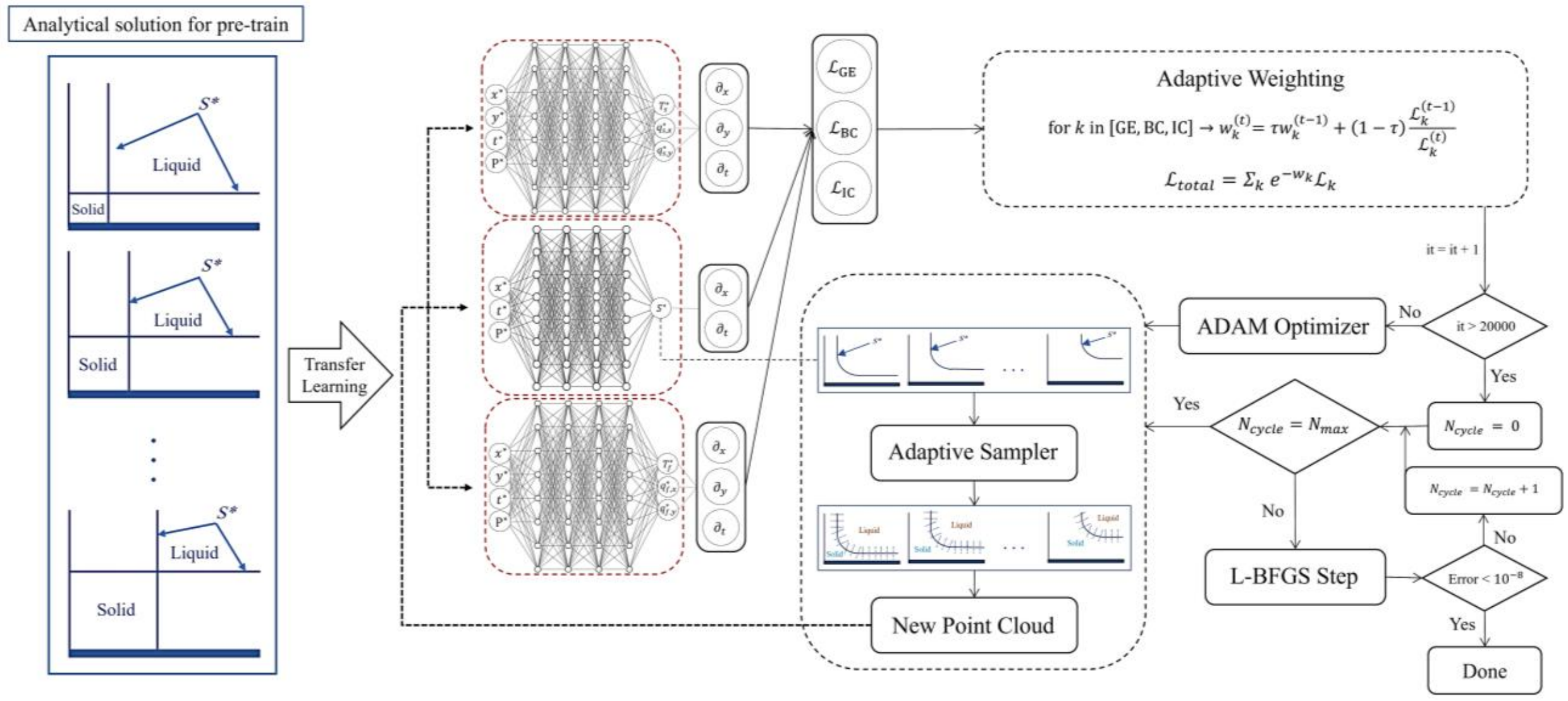

**Figure 2.** The architecture of the proposed Physics-Driven Deep learning (PDDL) framework introduced in this study.

---

**Algorithm 1** Adaptive PDDL for Two-Phase Stefan Problem

---

**Pre-training**
Prepare analytical solution: $\{(x_i, t_i, s_i^{target}, T_{f_i}^{target})\}$
Train networks to minimize: $\sum(s - s^{target})^2 + (T_f - T_f^{target})^2$
Train for $N_{pre}$ epochs with Adam ($lr = 5 \times 10^{-4}$)
**Phase 1: Adam Optimization with Adaptive Sampling**
**for** $i = 1$ to $N_{Adam}$ **do**
    $S \leftarrow \text{Net}_2(\mathcal{P}_{col}^S)$ ▷ Predict interface
    $\mathcal{P}_{enhanced} \leftarrow \text{AdaptiveSampling}(S, \mathcal{P}_{col})$
    Compute losses: $\mathcal{L}_{bc}, \mathcal{L}_{ic}, \mathcal{L}_{int}, \mathcal{L}_{pde}$
    Update weights: $w_k \leftarrow 0.8w_k + 0.2\frac{\mathcal{L}_k}{\mathcal{L}_k^{prev}}$
    $\mathcal{L}_{total} \leftarrow \sum e^{-w_k}\mathcal{L}_k$
    Update networks: $\theta \leftarrow \theta - \eta\nabla_\theta\mathcal{L}_{total}$
**end for**
**Phase 2: Adaptive L-BFGS Refinement**
**for** cycle $= 1$ to $N_{cycles}$ **do**
    $S \leftarrow \text{Net}_2(\mathcal{P}_{col}^S)$
    $\mathcal{P}_{fixed} \leftarrow \text{AdaptiveSampling}(S, \mathcal{P}_{col})$
    **for** $j = 1$ to $N_{LBFGS}$ **do**
        Compute $\mathcal{L}_{total}$ on $\mathcal{P}_{fixed}$
        L-BFGS step with strong Wolfe line search
        **if** $\|\Delta\mathcal{L}\| < 10^{-8}$ **then break**
        **end if**
    **end for**
**end for**
**Final Refinement**
$\mathcal{P}_{final} \leftarrow \text{AdaptiveSampling}(S_{final}, \mathcal{P}_{col})$
L-BFGS optimization on $\mathcal{P}_{final}$
**Output:** Trained networks and interface evolution

---

To incorporate the governing equations, initial conditions, and boundary conditions into the proposed methodology, the equations presented in the previous section have been non-dimensionalized. This procedure is adopted because non-dimensionalization balances the various terms in the loss function, thereby improving and stabilizing the learning process. For this purpose, the dimensionless variables are defined according to Equation (15).

$$x^* = \frac{x}{l_c}, \quad y^* = \frac{y}{l_c}, \quad S^* = \frac{S}{l_c}, \quad \mathrm{P}^* = \frac{\mathrm{P} - \mathrm{P}_{min}}{\mathrm{P}_{max} - \mathrm{P}_{min}}, \quad t^* = \frac{k_s t}{\rho_s c_s {l_c}^2}$$
$$T_s^* = \frac{T_s - T_\infty}{T_m - T_\infty}, \quad T_f^* = \frac{T_f - T_\infty}{T_m - T_\infty}, \quad q_s^* = \frac{\frac{\partial T_s}{\partial x}}{\frac{T_m - T_\infty}{l_c}}, \quad q_f^* = \frac{\frac{\partial T_f}{\partial x}}{\frac{T_m - T_\infty}{l_c}} \tag{15}$$

Using the introduced dimensionless variables, the equations from the prior section can be expressed as shown in Equations (16) to (33). A critical point to note is that during the network training, the right-hand side of these equations (the physics-informed residuals) must converge towards zero. Furthermore, as evident from the formulations, the computational time is reduced by defining the Fourier heat flux. This transformation allows terms originally involving higher-order derivatives to be reformulated in terms of first-order derivatives, enhancing computational time.

$$\frac{\partial T_s^*}{\partial t^*} - \left(\frac{\partial q_{s,x}^*}{\partial x^*} + \frac{\partial q_{s,y}^*}{\partial y^*}\right) = r_{\mathrm{GE},1} \tag{16}$$

$$q_{s,x}^* - \frac{\partial T_s^*}{\partial x^*} = r_{\mathrm{GE},2} \tag{17}$$

$$q_{s,y}^* - \frac{\partial T_s^*}{\partial y^*} = r_{\mathrm{GE},3} \tag{18}$$

$$\alpha_{s,f}\left(\frac{\partial T_f^*}{\partial t^*}\right) - \left(\frac{\partial q_{f,x}^*}{\partial x^*} + \frac{\partial q_{f,y}^*}{\partial y^*}\right) = r_{\mathrm{GE},4} \tag{19}$$

$$q_{f,x}^* - \frac{\partial T_f^*}{\partial x^*} = r_{\mathrm{GE},5} \tag{20}$$

$$q_{f,y}^* - \frac{\partial T_f^*}{\partial y^*} = r_{\mathrm{GE},6} \tag{21}$$

$$T_s^*(x^*, y^*, 0) - 1 = r_{\mathrm{IC},1} \tag{22}$$

$$T_f^*(x^*, y^*, 0) - 1 = r_{\mathrm{IC},2} \tag{23}$$

$$S^*(x^*, 0) = r_{\mathrm{IC},3} \tag{24}$$

$$\frac{\partial T_s^*(0, y^*, t^*)}{\partial x^*} - \mathrm{Bi}_s T_s^*(0, y^*, t^*) = r_{\mathrm{BC},1} \tag{25}$$

$$\frac{\partial T_f^*(0, y^*, t^*)}{\partial x^*} - \mathrm{Bi}_f T_f^*(0, y^*, t^*) = r_{\mathrm{BC},2} \tag{26}$$

$$T_s^*(x^*, \delta^*, t^*) - T_f^*(x^*, \delta^*, t^*) = r_{\mathrm{BC},3} \tag{27}$$

$$\frac{\partial T_s^*}{\partial x^*}\left(\frac{l_f}{2l_c}, y^*, t^*\right) = r_{\mathrm{BC},4} \tag{28}$$

$$\frac{\partial T_s^*}{\partial y^*}(x^*, 1, t^*) = r_{\mathrm{BC},5} \tag{29}$$

$$\frac{\partial T_f^*}{\partial x^*}\left(\frac{l_f}{2l_c}, y^*, t^*\right) = r_{\mathrm{BC},6} \tag{30}$$

$$\frac{\partial T_f^*}{\partial y^*}(x^*, 0, t^*) = r_{\mathrm{BC},7} \tag{31}$$

$$T_s^*(x^*, S^*(x^*, t^*), t^*) - 1 = r_{\mathrm{BC},8} \tag{32}$$

$$q_{s,y}^*\left(1 + \left(\frac{\partial S^*}{\partial x^*}\right)^2\right) - \frac{1}{\mathrm{Ja}}\frac{\partial S^*}{\partial t^*} = r_{\mathrm{BC},9} \tag{33}$$

In the aforementioned equations, Ja represents the Jakob number, $\mathrm{Bi}_s$ denotes the solid PCM Biot number, $\mathrm{Bi}_f$ is the fin Biot number, and $\alpha_{s,f}$ signifies the ratio of the thermal diffusivity of the solid phase to that of the fin, as defined by Equation (34).

$$\mathrm{Ja} = \frac{c_s(T_m - T_\infty)}{L}, \qquad \mathrm{Bi}_s = \frac{hl_c}{k_s}, \qquad \mathrm{Bi}_f = \frac{hl_c}{k_f}, \qquad \alpha_{s,f} = \frac{k_s \rho_f c_f}{\rho_s c_s k_f} \tag{34}$$

Based on the aforementioned equations, the loss function for the neural network illustrated in Figure 2 can be formulated as presented in Equation (35). In this equation, $N_{\mathrm{GE}}$, $N_{\mathrm{IC}}$, and $N_{\mathrm{BC}}$ denote the number of collocation points sampled from the governing equations, initial conditions, and boundary conditions, respectively. A key aspect of the adaptive sampling strategy employed in this study is that $N$ is not a fixed hyperparameter; rather, its value is dynamically adjusted by the algorithm to preferentially reduce the loss in the vicinity of the moving boundary, thereby enhancing predictive accuracy around this critical region.

$$\mathcal{L}_{\text{total}} = \sum_{i=1}^{6} \frac{1}{N_{\text{GE},i}} \sum_{j=1}^{N_{\text{GE},i}} r_{\text{GE},i}^2 + \sum_{i=1}^{3} \frac{1}{N_{\text{IC},i}} \sum_{j=1}^{N_{\text{IC},i}} r_{\text{IC},i}^2 + \sum_{i=1}^{9} \frac{1}{N_{\text{BC},i}} \sum_{j=1}^{N_{\text{BC},i}} r_{\text{BC},i}^2 \tag{35}$$

The methodology section previously introduced the strategy used for selecting collocation points within the solution domain; this approach is elaborated here. As illustrated in Figure 3, the point distribution is generated using the Latin Hypercube Sampling (LHS) technique. Figure 3(a) presents the distribution that would be obtained for a non-parametric PINN, with a fixed $\mathrm{P}^* = 1$. However, for the parametric PDDL employed in this study, the distribution must be modified to account for the geometric parameter $\mathrm{P}^*$, defined as the aspect ratio between $l_c$ and $l_f$. As shown in Figure 3(b), variations in the right-side symmetry and horizontal boundaries result in a range of admissible point locations between $x^* = 0.5$ and $x^* = 2.5$. In this figure, all point clouds generated at each time step are visualized together to illustrate the full span of sampled domains. Moreover, Figure 3(c) highlights the distinction between the initially generated background points and the additional collocation points introduced dynamically by the adaptive refinement algorithm during training. This adaptive mechanism effectively concentrates points near the evolving interface, which is essential for accurately resolving its motion and capturing the underlying physics. Algorithm 2 details the curvature-aware adaptive sampling procedure that dynamically refines collocation points near the moving interface.

(a)

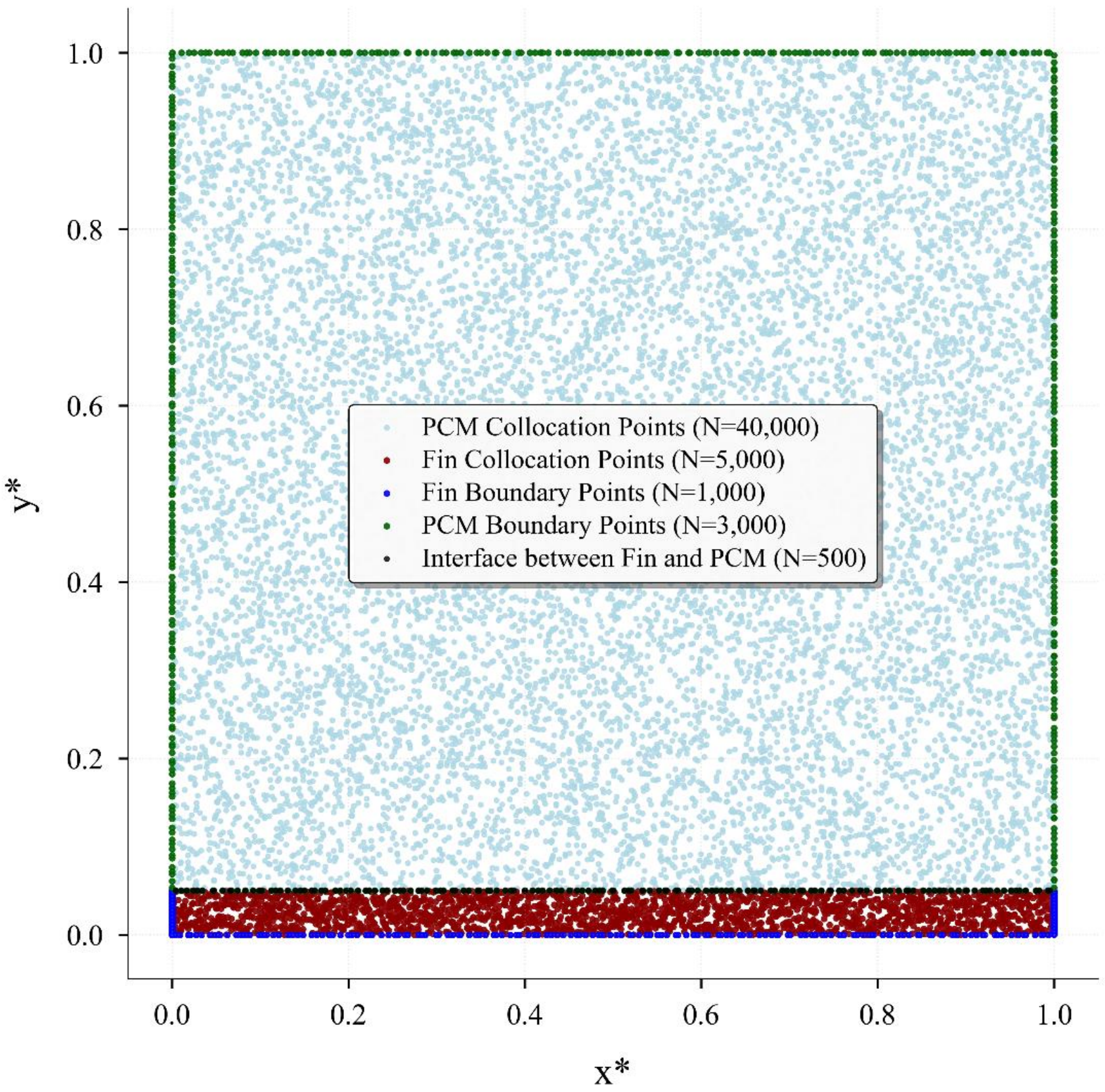

1.0
0.8
0.6
0.4
0.2
0.0
y*
PCM Collocation Points (N=40,000)
Fin Collocation Points (N=5,000)
Fin Boundary Points (N=1,000)
PCM Boundary Points (N=3,000)
Interface between Fin and PCM (N=500)
0.0
0.2
0.4
0.6
0.8
1.0
x*

(b)

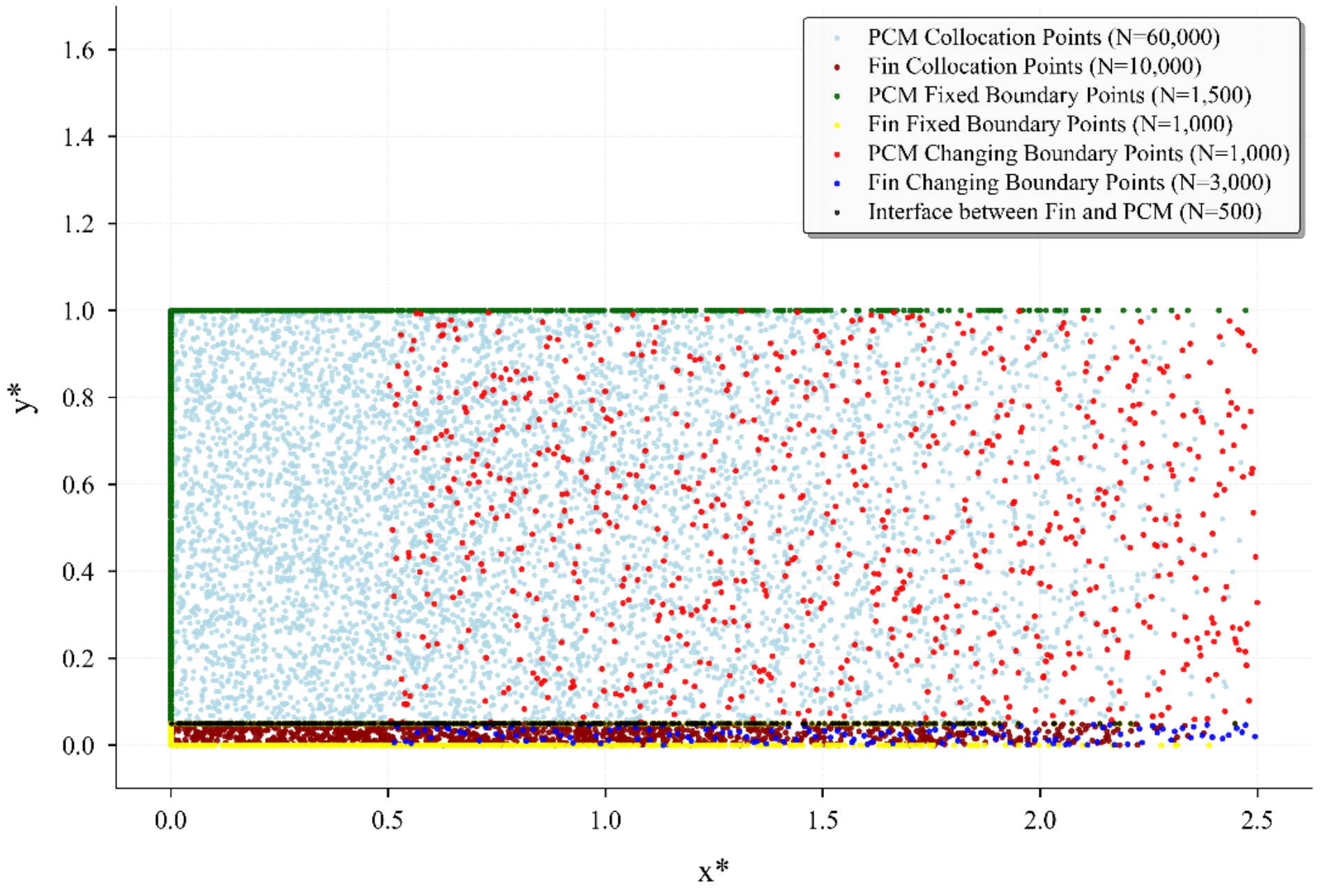

1.6
1.4
1.2
1.0
0.8
0.6
0.4
0.2
0.0
y*
PCM Collocation Points (N=60,000)
Fin Collocation Points (N=10,000)
PCM Fixed Boundary Points (N=1,500)
Fin Fixed Boundary Points (N=1,000)
PCM Changing Boundary Points (N=1,000)
Fin Changing Boundary Points (N=3,000)
Interface between Fin and PCM (N=500)
0.0
0.5
1.0
1.5
2.0
2.5
x*

(c)

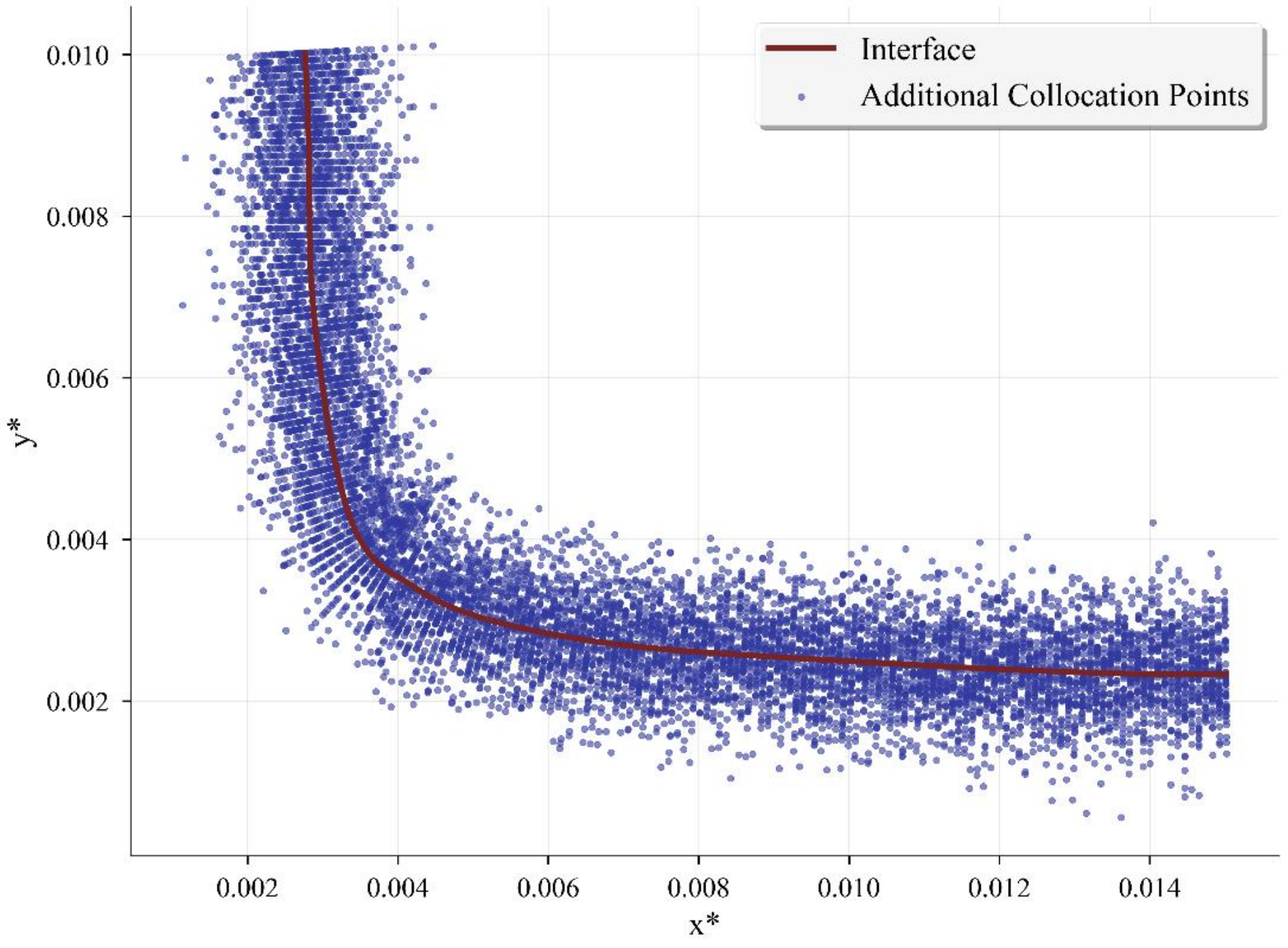

**Figure 3.** Domain point selection using Latin Hypercube Sampling (LHS): (a) Initial training point distribution; (b) Fixed and adaptive points during the adaptive sampling process; (c) Point distribution near the moving interface using adaptive sampling.

---

**Algorithm 2** Adaptive Sampling Procedure

---

**procedure** ADAPTIVESAMPLING($S, \mathcal{P}_{col}$)
  Sort points by $x$-coordinate
  Compute interface normals: $\mathbf{n} = (-s_x, 1)/\sqrt{1 + s_x^2}$
  Calculate curvature: $\kappa = |s_{xx}|/(1 + s_x^2)^{3/2}$
  Sample $N_{extra} = 500$ points using curvature weights
  Generate points: $\mathbf{x}_{extra} = \mathbf{x} + \sigma\xi\mathbf{n}$, $\xi \sim \mathcal{N}(0, 1)$
  Separate domains:
    $\mathcal{P}_{solid} = \{\mathbf{x} \mid y \leq s(\mathbf{x}) - 0.001\}$
    $\mathcal{P}_{liquid} = \{\mathbf{x} \mid y \geq s(\mathbf{x}) + 0.001\}$
    $\mathcal{P}_{fluid} = \mathcal{P}_{col}^{f}$
  **return** Enhanced point cloud
**end procedure**

---

## 4. Results

This section presents the simulation results obtained using the methodology introduced in the Methodology section. For validation, the results from the PDDL approach were compared with those from the studies by Mosaffa et al. [18] and Kothari et al. [39]. Figure 4 compares the location of the solid-liquid moving boundary within the PCM at two different times with one-dimensional analytical solutions. As can be observed, the PDDL method successfully predicted the interface location with high accuracy, effectively capturing the growth of the solid layer within the PCM over time from Figure 4(a) to 4(b). Furthermore, given the one-dimensional nature of the modeling in the referenced studies, some discrepancies between the results of the present method and theirs are to be expected.

(a)

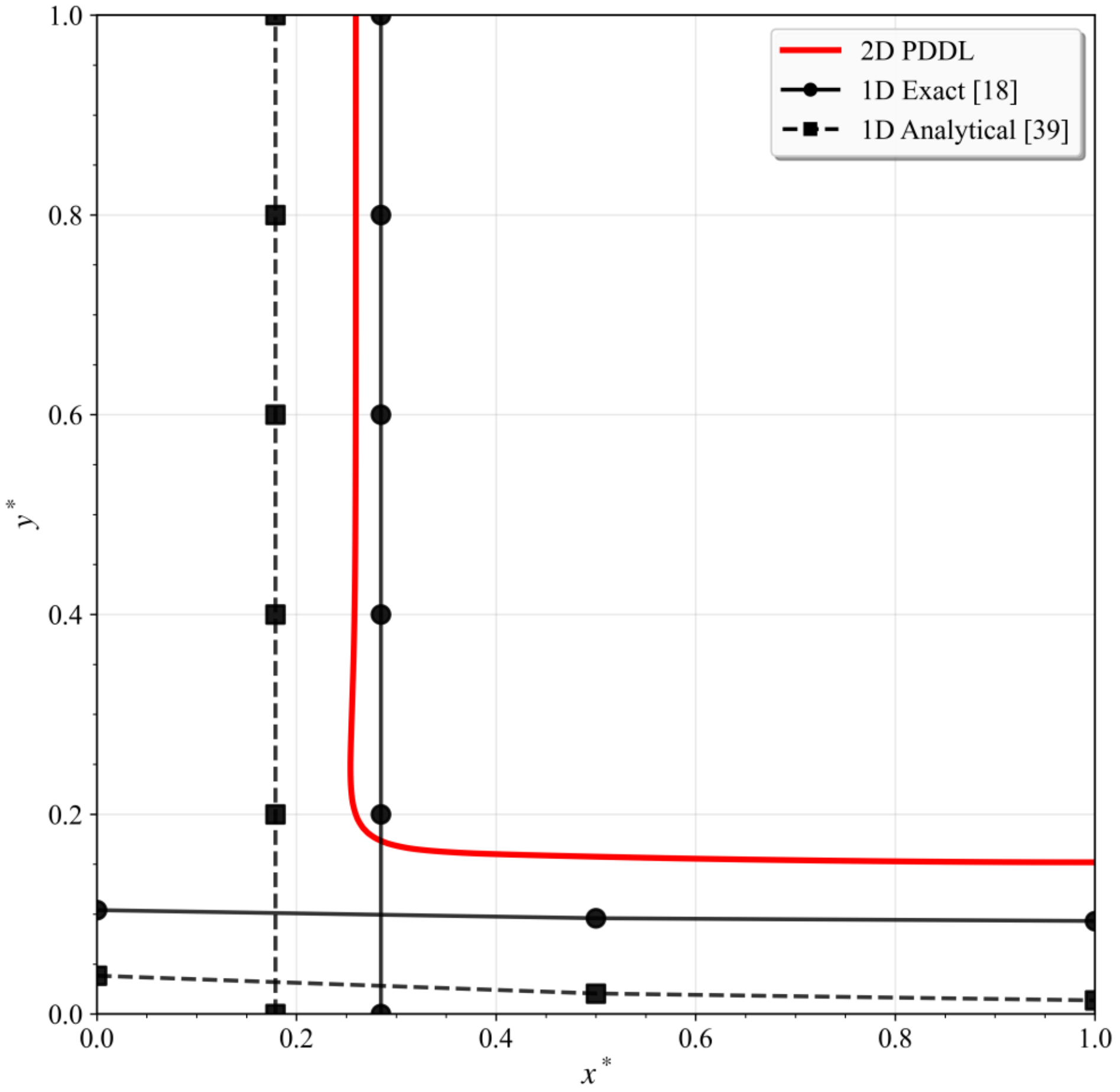

(b)

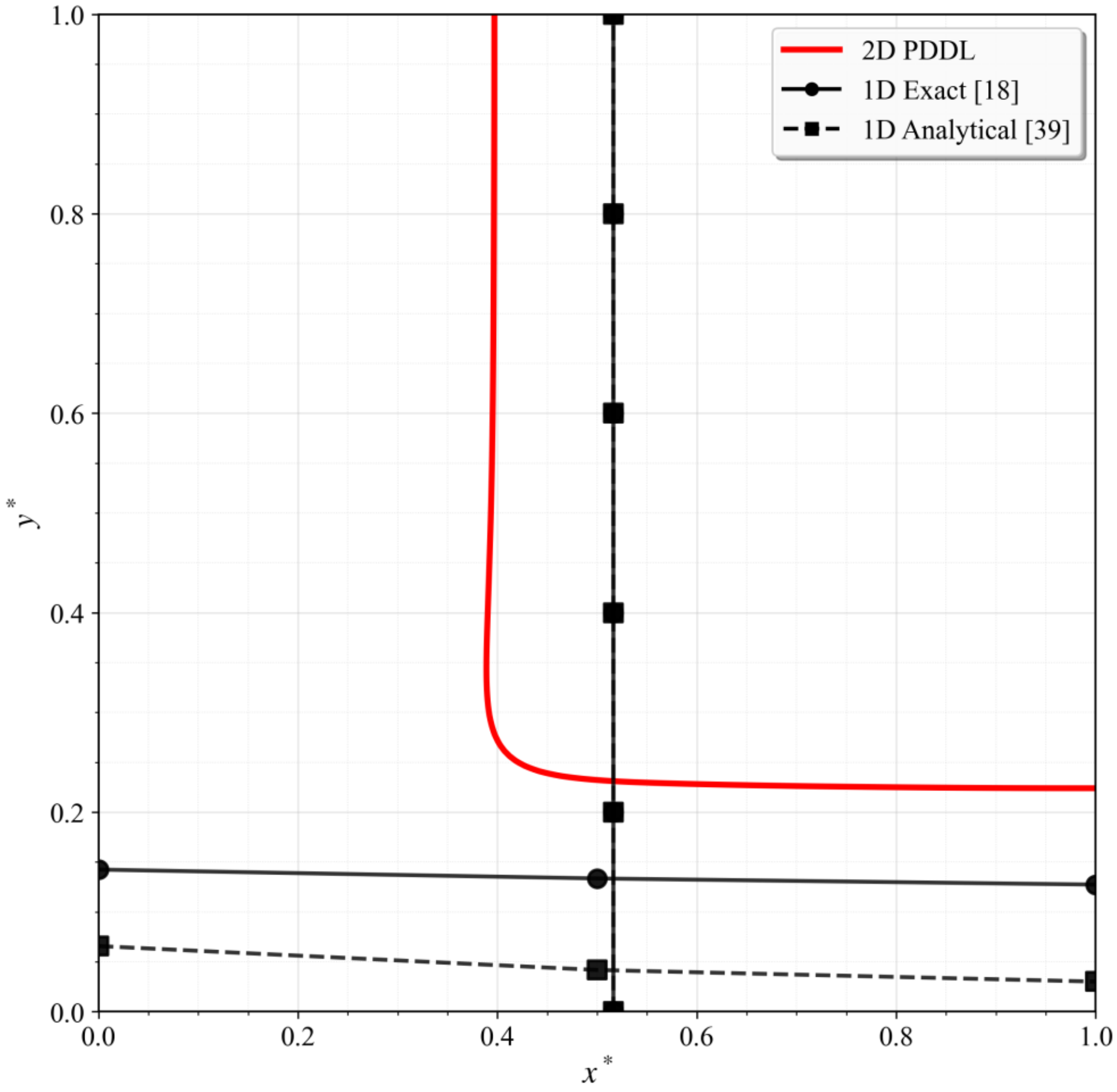

**Figure 4.** Comparison of the solid-liquid interface position within the PCM obtained from the present study with one-dimensional solutions from Mosaffa et al. [18] and Kothari et al. [39] for a geometric parameter $P^* = 2$. (a) $t^* = 1.61$; (b) $t^* = 3.23$.

Subsequently, the fin temperature was examined as a metric to evaluate the proposed solution against the one-dimensional results. As observed in Figure 5, the fin temperature calculated by the proposed method exhibits satisfactory agreement with the temperature profile derived from the one-dimensional approximation of the problem. As the figure indicates, although the proposed two-dimensional solution follows the general trend of the one-dimensional fin temperature variation, a maximum error of 0.4% is observable in the results. This error margin can be attributed to the high thermal conductivity of the fin material, which results in negligible temperature variation along the fin's length. Over time, heat transfer from the PCM to the surrounding air causes the PCM temperature to decrease.

Consequently, the fin temperature, which facilitates this heat transfer process, also decreases with time, as illustrated in the figure.

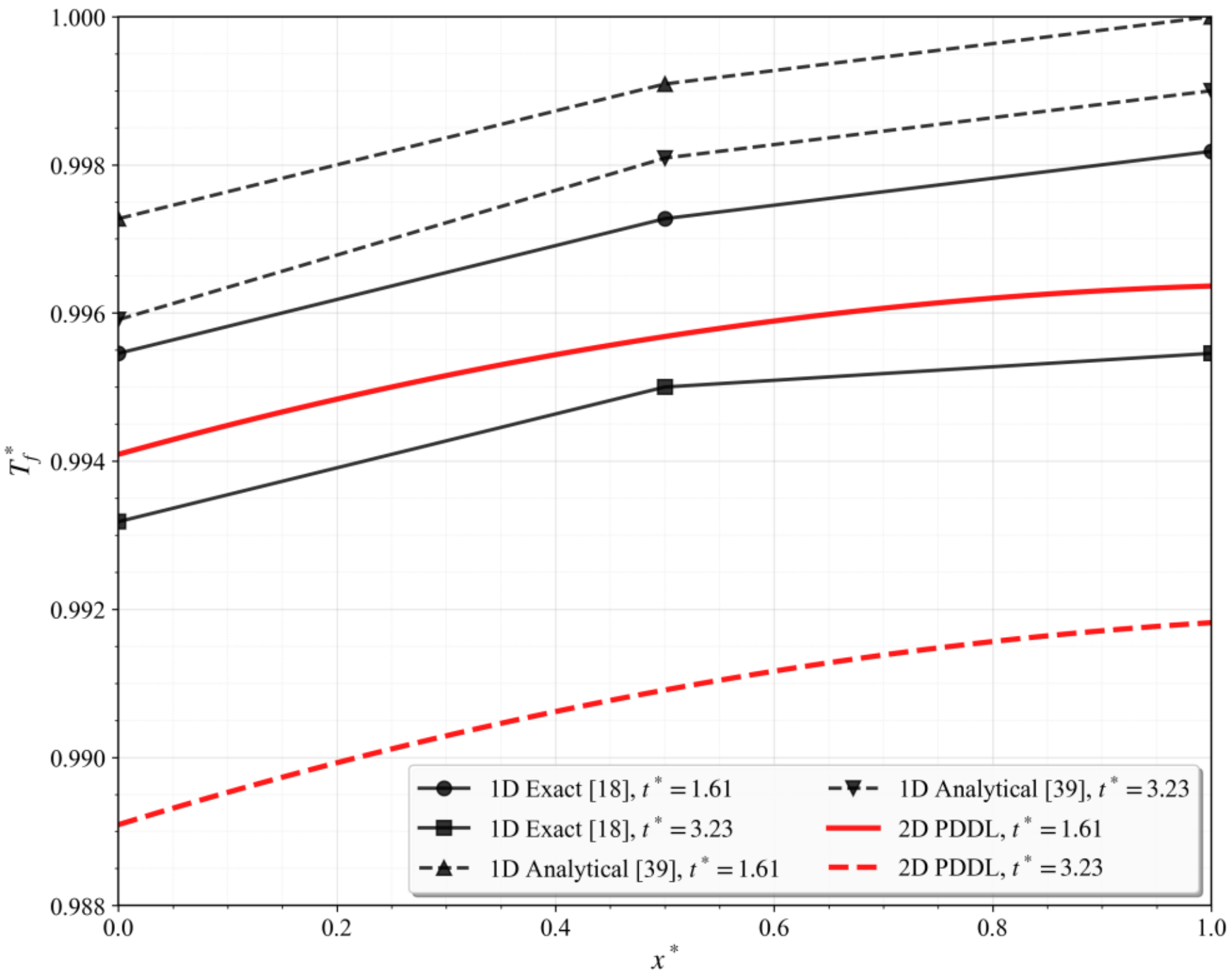

**Figure 5.** Comparison of the fin temperature obtained by the present method with the results from the studies by Mosaffa et al. [18] and Kothari et al. [39] for a geometric parameter $P^* = 2$.

The solid fraction within the PCM, which indicates the proportion of the storage unit that has solidified at any given moment, served as another metric for comparing the results of the proposed method with the one-dimensional benchmark solution. Figure 6 compares the temporal evolution of the solidification fraction obtained from the present study with the results from the one-dimensional reference study. As the figure demonstrates, the proposed method successfully predicts the solidification history, showing strong agreement between the PDDL results and the one-dimensional solution in this regard.

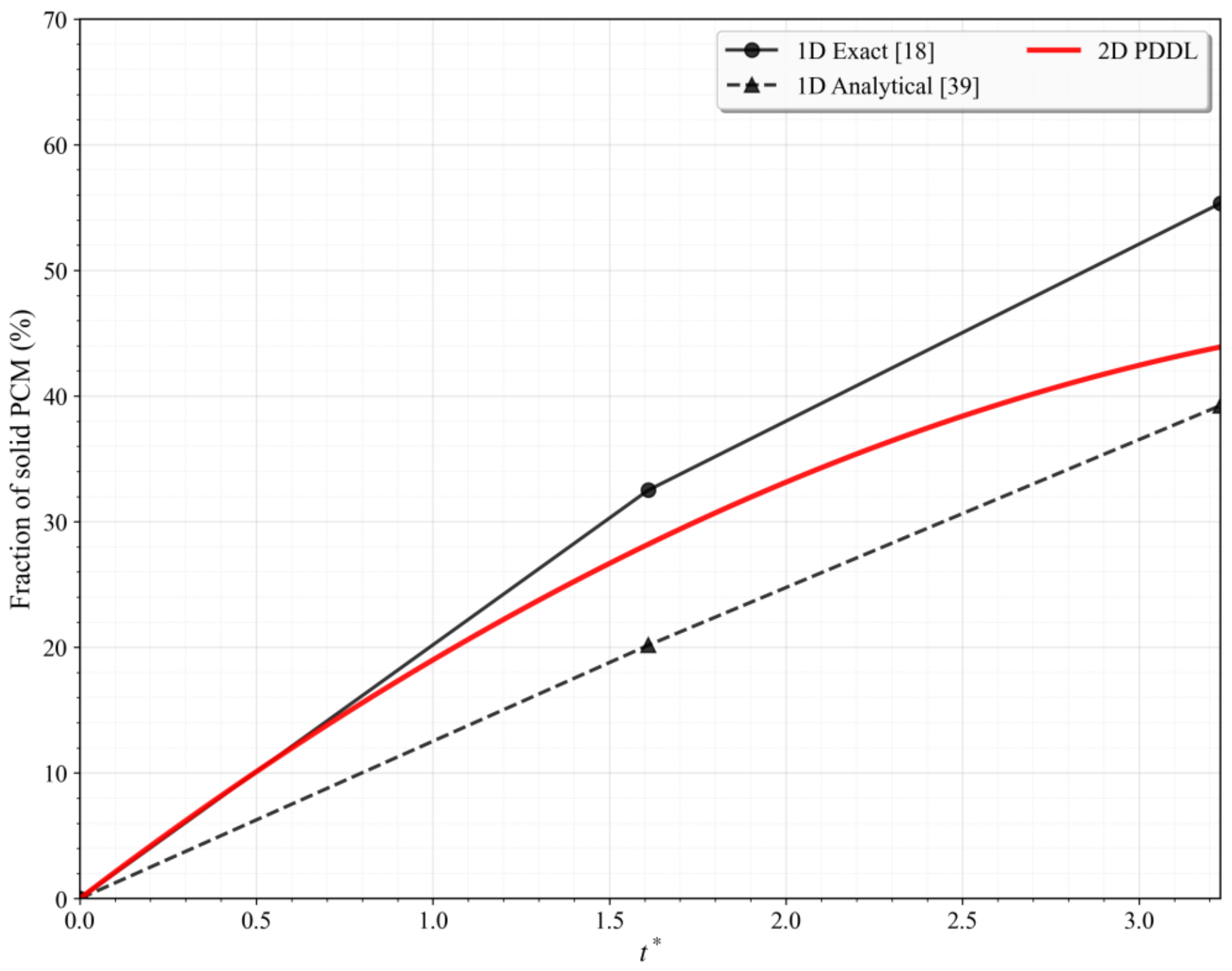

**Figure 6.** Temporal evolution of the solid fraction in the PCM predicted by the PDDL method, compared against one-dimensional solutions, for a geometric parameter $P^* = 2$.

Based on the parametric study conducted in this research, the contours of dimensionless temperature and the location of the solid-liquid interface for various values of the parameter $P^*$ are presented in Figure 7. By incorporating the $P^*$ parameter as an input to the neural network, the proposed method is capable of generating predictions across the parameter range after a single training process. Furthermore, to maintain consistency with the validated results, the contours are displayed at the same two normalized time instances as before. The complete set of results obtained using the PDDL method is available in the GitHub repository associated with this study; here, we confine our presentation to a representative subset of results within the $t^*$ and $P^*$ ranges. As illustrated in the figure, the progression of solidification is evident across all $P^*$ values, with a larger portion of the PCM transitioning to the solid phase as normalized time increases. Furthermore, at a specific time instant, a decrease in the $P^*$ parameter, which corresponds to a lower aspect ratio, accelerates the solidification process. Consequently, a larger volume fraction of the PCM is occupied by the solid phase.

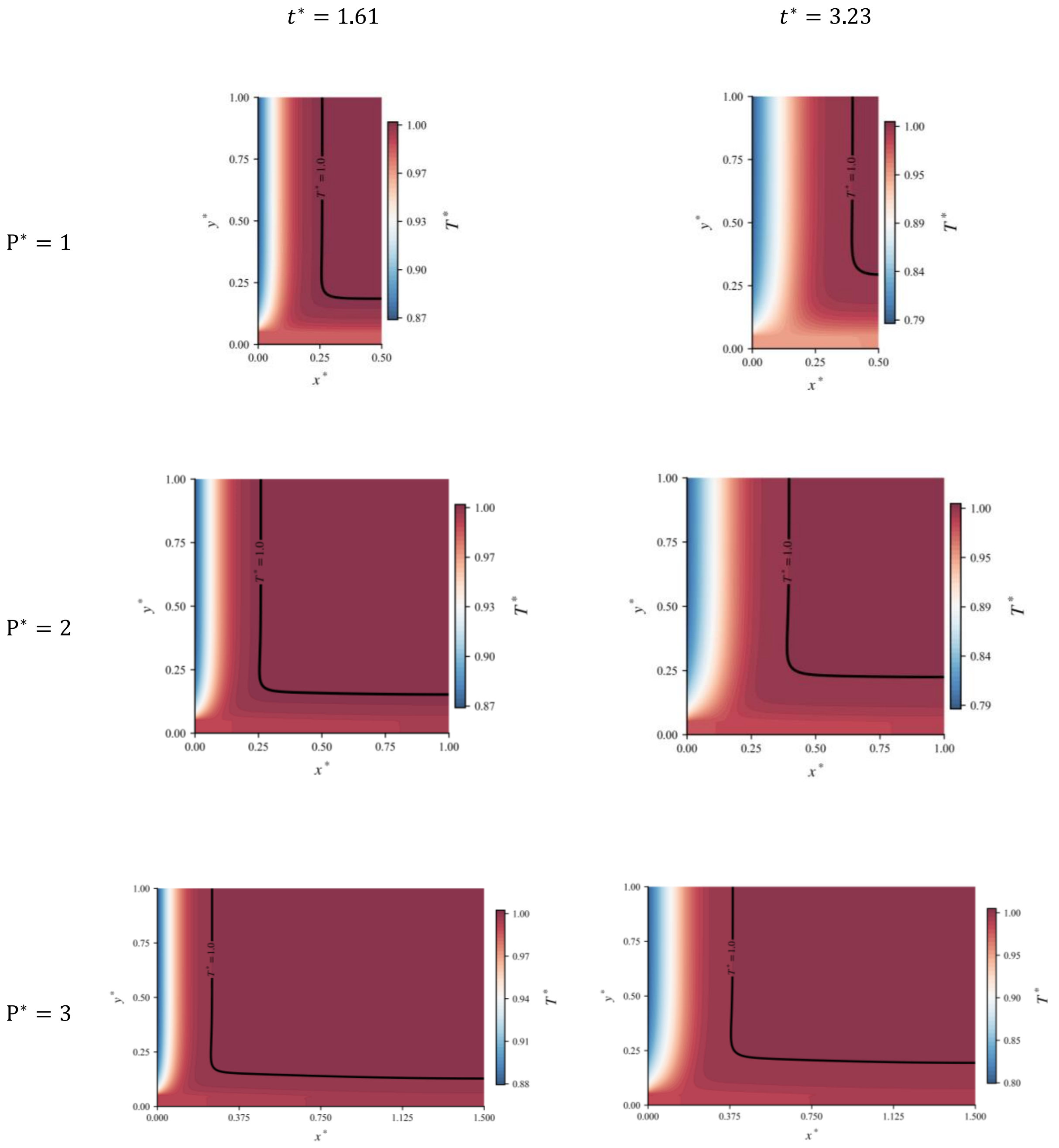

**Figure 7.** Contours of the dimensionless temperature and the solid-liquid interface position predicted by the PDDL method.

For a more detailed analysis of the results obtained by the PDDL-based solution, Figure 8 illustrates the temporal evolution of the solid fraction for different values of the geometric parameter $P^*$. At the maximum normalized time shown on the horizontal axis, which is determined by the specified range for the normalized temperature in the PDDL framework, an increase in the $P^*$ parameter leads to a decrease in the solid fraction of the PCM. This observation indicates that a larger $P^*$value, representing a higher aspect ratio and consequently a larger initial volume of liquid PCM, requires a longer time to complete the solidification process.

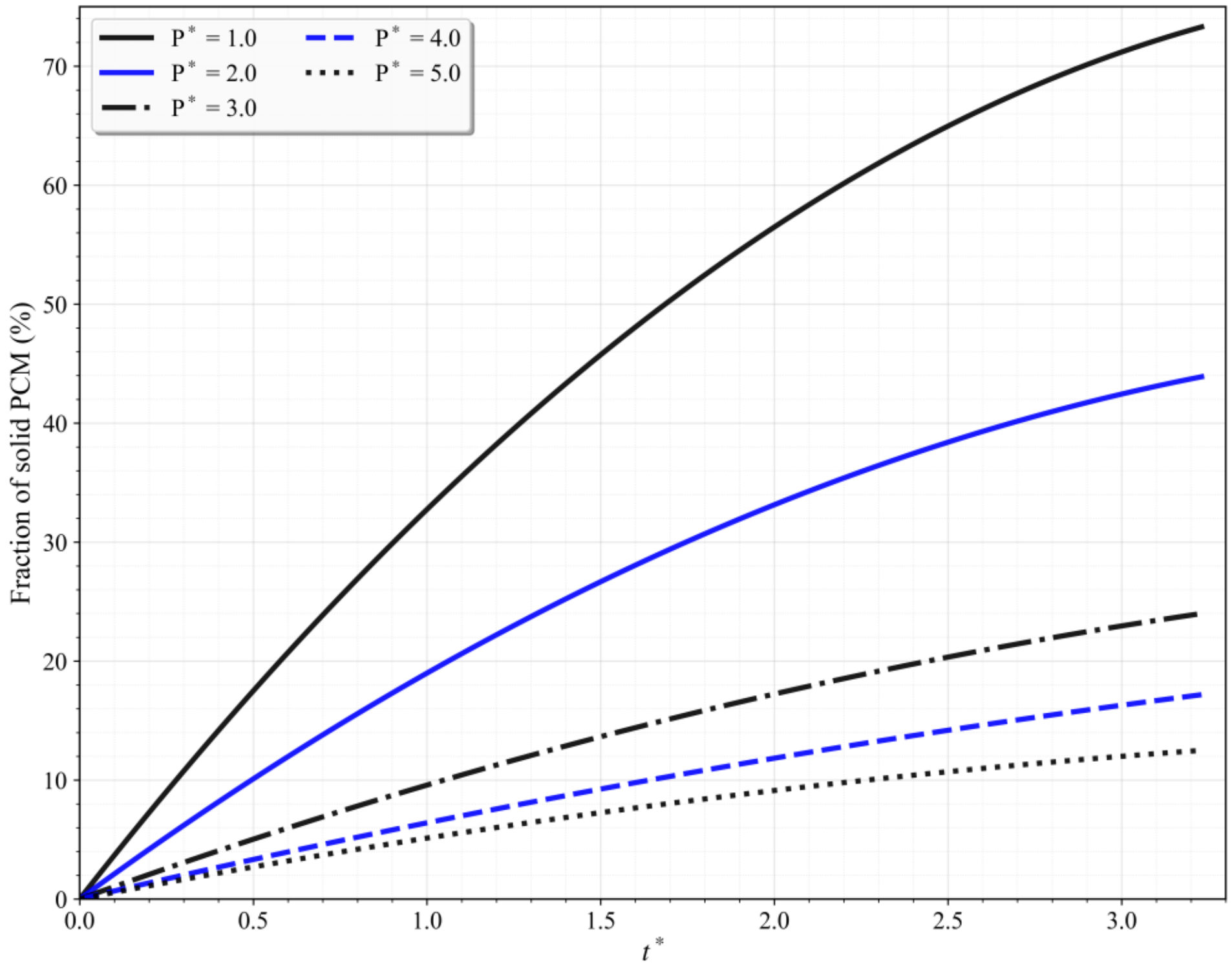

**Figure 8.** Temporal evolution of the solid fraction for different values of the geometric parameter $P^*$.

In the final part of the study, the maximum and minimum fin temperatures at the end of the solidification process were examined for different values of the parameter $P^*$. As shown in Figure 9, an increases in the $P^*$ parameter, which corresponds to a higher aspect ratio, results in a higher maximum temperature. Furthermore, it can be observed that as $P^*$ increases, the difference between the minimum and maximum fin temperatures also increases, reaching its peak value at $P^* = 4.7$.

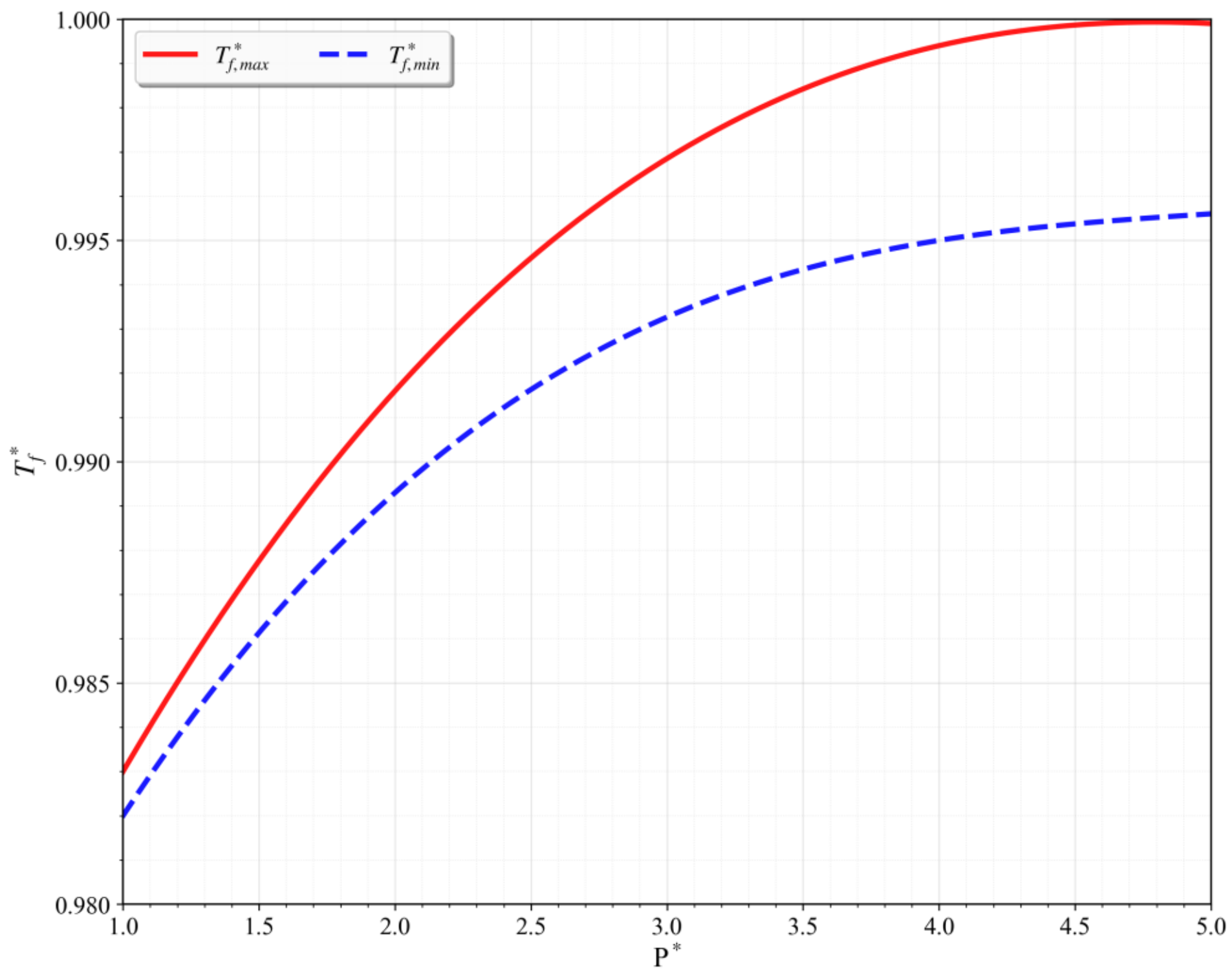

**Figure 9.** Minimum and maximum fin temperatures at the end of the solidification process for various values of the geometric parameter $P^*$.

## Conclusion

This study successfully developed and validated a Physics-Driven Deep Learning (PDDL) framework for simulating the complex Moving Boundary Problem (MBP) in phase change material solidification with fin enhancement. The proposed multi-network architecture demonstrated exceptional capability in simultaneously predicting the solid-liquid interface evolution, fin temperature distribution, and temporal solid fraction progression across various geometric configurations. By incorporating physical constraints directly into the learning process through adaptive sampling and loss balancing, the method effectively captured the intricate dynamics of the phase transition process under convective boundary conditions. The parametric formulation enabled comprehensive analysis of geometric influences on solidification behavior, providing valuable insights for thermal energy storage system optimization. Quantitative validation against established one-dimensional analytical solutions confirmed the framework's accuracy and reliability in handling moving boundary challenges without requiring mesh regeneration. The method's mesh-free nature, coupled with its ability to learn from physical principles rather than extensive labeled data, represents a significant advancement over conventional numerical approaches. These findings highlight the strong potential of PDDL methods for complex multiphysics problems involving interface tracking. The framework's architecture readily extends to three-dimensional configurations, multi-material systems, and other applications where precise interface prediction is critical, such as crystal growth, biomedical engineering, and advanced manufacturing processes. Future work will focus on experimental validation and extension to more complex boundary conditions and material systems.

## CRediT authorship contribution statement

**Meraj Hassanzadeh:** Conceptualization, Methodology, Validation, Visualization, Software, Investigation, Writing the original draft.

**Ehsan Ghaderi:** Conceptualization, Methodology, Validation, Visualization, Software, Investigation, Writing the original draft.

**Fatemeh Fatahi:** Methodology, Visualization, Review and Editing.

**Mohamad Ali Bijarchi:** Review and Editing, Supervision.

## Acknowledgment

The authors gratefully acknowledge the assistance of AI-based language tools in improving the clarity and fluency of the English writing. All ideas, arguments, and scientific content are solely the responsibility of the authors.

## Funding

This research received no specific grant from any funding agency.

## Conflict of Interest

The authors declare no conflict of interest.

## References

[1] Douvi, E., Pagkalos, C., Dogkas, G., Koukou, M.K., Stathopoulos, V.N., Caouris, Y. and Vrachopoulos, M.G., 2021. Phase change materials in solar domestic hot water systems: A review. *International Journal of Thermofluids*, 10, p.100075. https://doi.org/10.1016/j.ijft.2021.100075.

[2] Mosaffa, A.H., Ferreira, C.I., Talati, F. and Rosen, M.A., 2013. Thermal performance of a multiple PCM thermal storage unit for free cooling. *Energy Conversion and Management*, *67*, pp.1-7. https://doi.org/10.1016/j.enconman.2012.10.018.

[3] Luo, J., Zou, D., Wang, Y., Wang, S. and Huang, L., 2022. Battery thermal management systems (BTMs) based on phase change material (PCM): A comprehensive review. *Chemical Engineering Journal*, *430*, p.132741. https://doi.org/10.1016/j.cej.2021.132741.

[4] Jain, A. and Parhizi, M., 2021. Conditionally exact closed-form solution for moving boundary problems in heat and mass transfer in the presence of advection. *International Journal of Heat and Mass Transfer*, *180*, p.121802. https://doi.org/10.1016/j.ijheatmasstransfer.2021.121802.

[5] Turkyilmazoglu, M., 2018. Stefan problems for moving phase change materials and multiple solutions. *International Journal of Thermal Sciences*, *126*, pp.67-73. https://doi.org/10.1016/j.ijthermalsci.2017.12.019.

[6] Mosaffa, A.H., Talati, F., Tabrizi, H.B. and Rosen, M.A., 2012. Analytical modeling of PCM solidification in a shell and tube finned thermal storage for air conditioning systems. *Energy and buildings*, *49*, pp.356-361. https://doi.org/10.1016/j.enbuild.2012.02.053.

[7] Talati, F. and Taghilou, M., 2015. Lattice Boltzmann application on the PCM solidification within a rectangular finned container. *Applied Thermal Engineering*, *83*, pp.108-120. https://doi.org/10.1016/j.applthermaleng.2015.03.017.

[8] Taghilou, M. and Talati, F., 2018. Analytical and numerical analysis of PCM solidification inside a rectangular finned container with time-dependent boundary condition. *International Journal of Thermal Sciences*, *133*, pp.69-81. https://doi.org/10.1016/j.ijthermalsci.2018.04.042.

[9] Lamberg, P., 2004. Approximate analytical model for two-phase solidification problem in a finned phase-change material storage. *Applied Energy*, *77*(2), pp.131-152. https://doi.org/10.1016/S0306-2619(03)00106-5.

[10] Liu, G., Xiao, T., Guo, J., Wei, P., Yang, X. and Hooman, K., 2022. Melting and solidification of phase change materials in metal foam filled thermal energy storage tank: Evaluation on gradient in pore structure. *Applied Thermal Engineering*, *212*, p.118564. https://doi.org/10.1016/j.applthermaleng.2022.118564.

[11] Masoumi, H. and Mirfendereski, S.M., 2022. Experimental and numerical investigation of melting/solidification of nano-enhanced phase change materials in shell & tube thermal energy storage systems. *Journal of Energy Storage*, *47*, p.103561. https://doi.org/10.1016/j.est.2021.103561.

[12] Xu, M., Hanawa, Y., Akhtar, S., Sakuma, A., Zhang, J., Yoshida, J., Sanada, M., Sasaki, Y. and Sasmito, A.P., 2023. Multi-scale analysis for solidification of phase change materials (PCMs): Experiments and modeling. *International Journal of Heat and Mass Transfer*, *212*, p.124182. https://doi.org/10.1016/j.ijheatmasstransfer.2023.124182.

[13] Zalba, B., Marín, J.M., Cabeza, L.F. and Mehling, H., 2003. Review on thermal energy storage with phase change: materials, heat transfer analysis and applications. *Applied thermal engineering*, *23*(3), pp.251-283. https://doi.org/10.1016/S1359-4311(02)00192-8.

[14] Zivkovic, B. and Fujii, I., 2001. An analysis of isothermal phase change of phase change material within rectangular and cylindrical containers. *Solar energy*, *70*(1), pp.51-61. https://doi.org/10.1016/S0038-092X(00)00112-2.

[15] Esen, M., 2000. Thermal performance of a solar-aided latent heat store used for space heating by heat pump. *Solar energy*, *69*(1), pp.15-25. https://doi.org/10.1016/S0038-092X(00)00015-3.

[16] Lamberg, P. and Siren, K., 2003. Approximate analytical model for solidification in a finite PCM storage with internal fins. *Applied Mathematical Modelling*, *27*(7), pp.491-513. https://doi.org/10.1016/S0307-904X(03)00080-5.

[17] Talati, F., Mosaffa, A.H. and Rosen, M.A., 2011. Analytical approximation for solidification processes in PCM storage with internal fins: imposed heat flux. *Heat and mass transfer*, *47*(4), pp.369-376. https://doi.org/10.1007/s00231-010-0729-9.

[18] Mosaffa, A.H., Talati, F., Rosen, M.A. and Tabrizi, H.B., 2012. Approximate analytical model for PCM solidification in a rectangular finned container with convective cooling boundaries. *International Communications in Heat and Mass Transfer*, *39*(2), pp.318-324. https://doi.org/10.1016/j.icheatmasstransfer.2011.11.015.

[19] Cofré-Toledo, J., Roa-Cossio, D., Vasco, D.A., Cabeza, L.F. and Rouault, F., 2022. Numerical simulation of the melting and solidification processes of two organic phase change materials in spherical enclosures for cold thermal energy storage applications. *Journal of Energy Storage*, *51*, p.104337. https://doi.org/10.1016/j.est.2022.104337.

[20] Rashid, F.L., Eisapour, M., Ibrahem, R.K., Talebizadehsardari, P., Hosseinzadeh, K., Abbas, M.H., Mohammed, H.I., Yvaz, A. and Chen, Z., 2023. Solidification enhancement of phase change materials using fins and nanoparticles in a triplex-tube thermal energy storage unit: Recent advances and development. *International Communications in Heat and Mass Transfer*, *147*, p.106922. https://doi.org/10.1016/j.icheatmasstransfer.2023.106922.

[21] Azad, M., Groulx, D. and Donaldson, A., 2023. Solidification of phase change materials in horizontal annuli. *Journal of Energy Storage*, *57*, p.106308. https://doi.org/10.1016/j.est.2022.106308.

[22] Mukhamediev, R.I., Symagulov, A., Kuchin, Y., Yakunin, K. and Yelis, M., 2021. From classical machine learning to deep neural networks: A simplified scientometric review. *Applied Sciences*, *11*(12), p.5541. https://doi.org/10.3390/app11125541.

[23] Yaghoubi, E., Yaghoubi, E., Khamees, A. and Vakili, A.H., 2024. A systematic review and meta-analysis of artificial neural network, machine learning, deep learning, and ensemble learning approaches in field of geotechnical engineering. *Neural Computing and Applications*, *36*(21), pp.12655-12699. https://doi.org/10.1007/s00521-024-09893-7.

[24] Jambunathan, K., Hartle, S.L., Ashforth-Frost, S. and Fontama, V.N., 1996. Evaluating convective heat transfer coefficients using neural networks. *International Journal of Heat and Mass Transfer*, *39*(11), pp.2329-2332. https://doi.org/10.1016/0017-9310(95)00332-0.

[25] Kim, J. and Lee, C., 2020. Prediction of turbulent heat transfer using convolutional neural networks. *Journal of Fluid Mechanics*, *882*, p.A18. https://doi.org/10.1017/jfm.2019.814.

[26] Di Natale, L., Svetozarevic, B., Heer, P. and Jones, C.N., 2022. Physically consistent neural networks for building thermal modeling: theory and analysis. *Applied Energy*, *325*, p.119806. https://doi.org/10.1016/j.apenergy.2022.119806.

[27] Cai, S., Wang, Z., Wang, S., Perdikaris, P. and Karniadakis, G.E., 2021. Physics-informed neural networks for heat transfer problems. *Journal of Heat Transfer*, *143*(6), p.060801. https://doi.org/10.1115/1.4050542.

[28] Cai, S., Mao, Z., Wang, Z., Yin, M. and Karniadakis, G.E., 2021. Physics-informed neural networks (PINNs) for fluid mechanics: A review. *Acta Mechanica Sinica*, *37*(12), pp.1727-1738. https://doi.org/10.1007/s10409-021-01148-1.

[29] Raissi, M., Perdikaris, P. and Karniadakis, G.E., 2019. Physics-informed neural networks: A deep learning framework for solving forward and inverse problems involving nonlinear partial differential equations. *Journal of Computational physics*, *378*, pp.686-707. https://doi.org/10.1016/j.jcp.2018.10.045.

[30] Raissi, M., Yazdani, A. and Karniadakis, G.E., 2020. Hidden fluid mechanics: Learning velocity and pressure fields from flow visualizations. *Science*, *367*(6481), pp.1026-1030. https://doi.org/10.1126/science.aaw4741.

[31] Cai, S., Wang, Z., Fuest, F., Jeon, Y.J., Gray, C. and Karniadakis, G.E., 2021. Flow over an espresso cup: inferring 3-D velocity and pressure fields from tomographic background oriented Schlieren via physics-informed neural networks. *Journal of Fluid Mechanics*, *915*, p.A102. https://doi.org/10.1017/jfm.2021.135.

[32] Liu, K., Luo, K., Cheng, Y., Liu, A., Li, H., Fan, J. and Balachandar, S., 2023. Surrogate modeling of parameterized multi-dimensional premixed combustion with physics-informed neural networks for rapid exploration of design space. *Combustion and Flame*, *258*, p.113094. https://doi.org/10.1016/j.combustflame.2023.113094.

[33] Cao, Z., Liu, K., Luo, K., Cheng, Y. and Fan, J., 2023. Efficient optimization design of flue deflectors through parametric surrogate modeling with physics-informed neural networks. *Physics of Fluids*, *35*(12). https://doi.org/10.1063/5.0180594.

[34] Lu, L., Pestourie, R., Yao, W., Wang, Z., Verdugo, F. and Johnson, S.G., 2021. Physics-informed neural networks with hard constraints for inverse design. *SIAM Journal on Scientific Computing*, *43*(6), pp.B1105-B1132. https://doi.org/10.1137/21M1397908.

[35] Ghaderi, E., Bijarchi, M., Hannani, S.K. and Nouri-Borujerdi, A., 2024. Parametric and inverse analysis of flow inside an obstructed channel under the influence of magnetic field using physics informed neural networks. *arXiv preprint arXiv:2404.17261*. https://doi.org/10.48550/arXiv.2404.17261.

[36] Priyadarshi, G., Murali, C., Agarwal, S. and Naik, B.K., 2024. Parametric investigation and optimization of phase change material-based thermal energy storage integrated desiccant coated energy exchanger through physics informed neural networks oriented deep learning approach. *Journal of Energy Storage*, *80*, p.110231. https://doi.org/10.1016/j.est.2023.110231.

[37] Kathane, S. and Karagadde, S., 2024. A Physics Informed Neural Network (PINN) Methodology for Coupled Moving Boundary PDEs. *arXiv preprint arXiv:2409.10910*. https://doi.org/10.48550/arXiv.2409.10910.

[38] Madir, B.E., Luddens, F., Lothodé, C. and Danaila, I., 2025. Physics Informed Neural Networks for heat conduction with phase change. *International Journal of Heat and Mass Transfer*, *252*, p.127430. https://doi.org/10.1016/j.ijheatmasstransfer.2025.127430.

[39] Kothari, R., Das, S., Sahu, S.K. and Kundalwal, S.I., 2019. Analysis of solidification in a finite PCM storage with internal fins by employing heat balance integral method. *International Journal of Energy Research*, *43*(12), pp.6366-6388. https://doi.org/10.1002/er.4363.

## Supplementary: One-Dimensional Problem Modeling

As mentioned in the main text, the problem can be modeled in one dimension to derive an analytical solution. Figure S1 (a) provides a schematic of the one-dimensional configuration. As shown, for the 1D modeling, the domain is divided into two distinct regions: in Region 1, a one-dimensional model of the solid phase is constructed, while in Region 2, a one-dimensional representation of the fin is achieved using the control volume approach, as illustrated. An important observation from Figure S1 (b) is that the two-phase region and the associated temperature variation within it are neglected in this model. Furthermore, temperature changes within the liquid phase are also disregarded, meaning all modes of heat transfer (conduction, convection, and radiation) within the liquid are omitted from the analysis. Figure S1 (c) illustrates the assumption of a linear temperature profile within the solid region, which is adopted to derive the fin temperature equation. This simplifying assumption is justifiable given the thin nature of the solid layer formed on the fin, making the linear approximation reasonably acceptable.

(a)

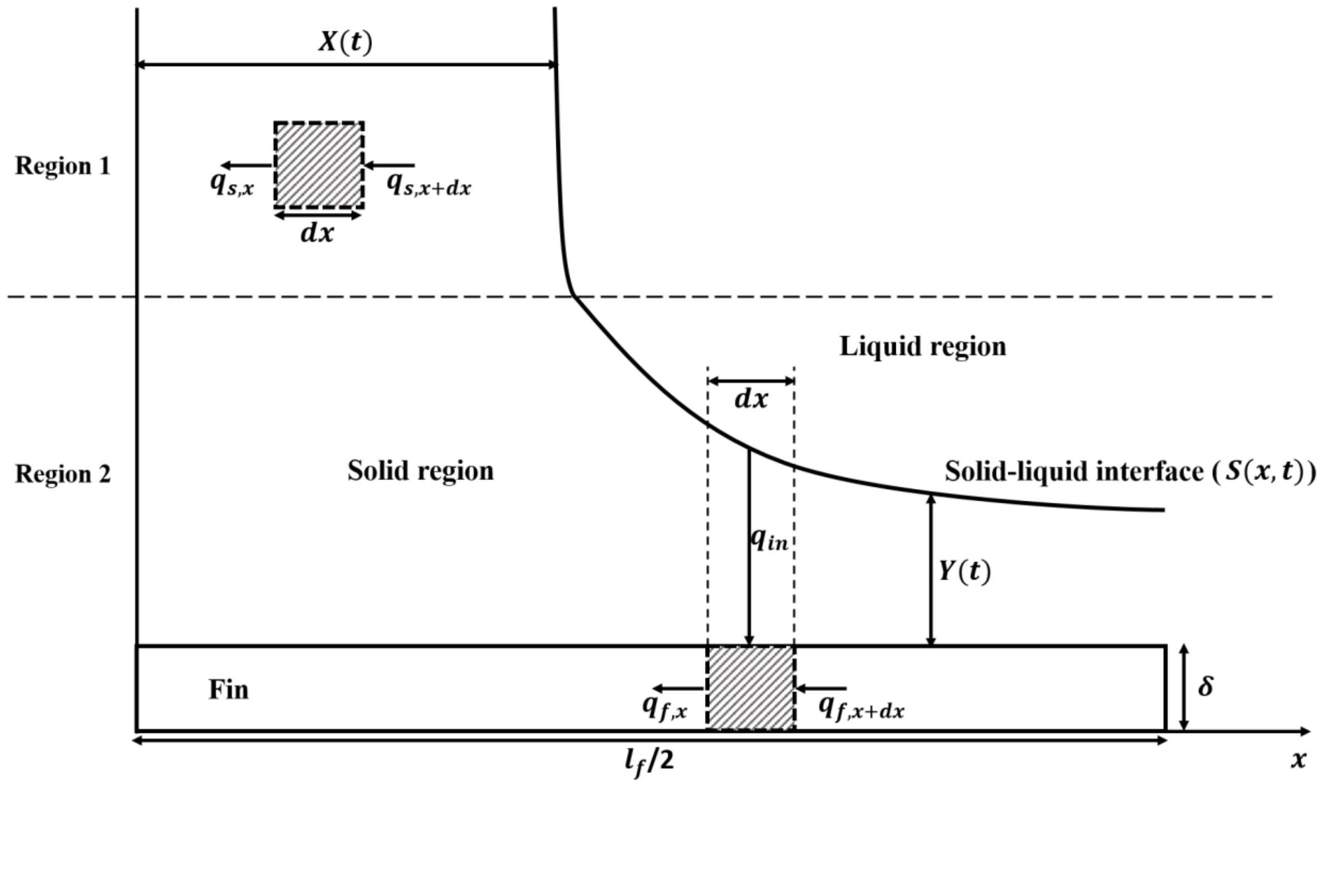

(b)

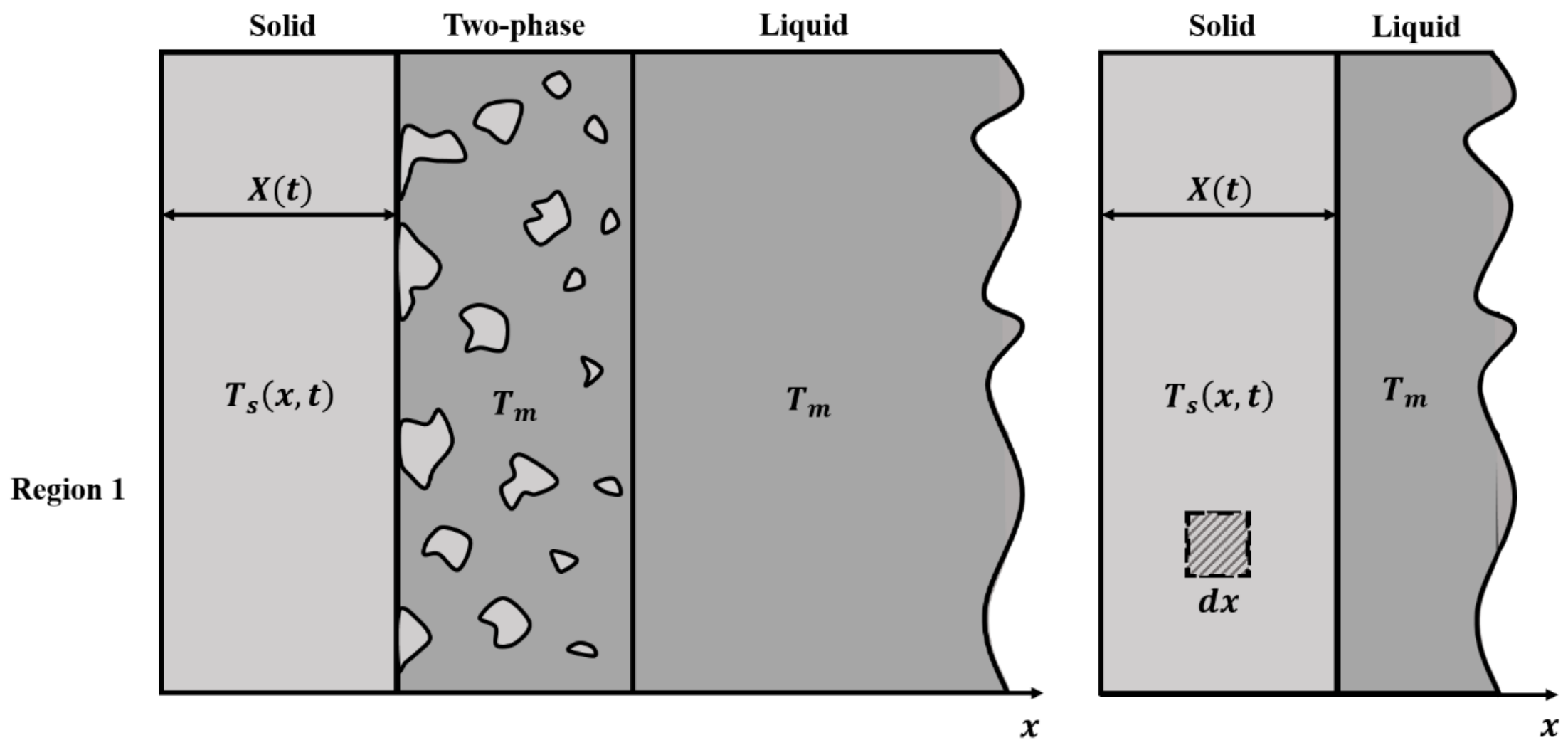

(c)

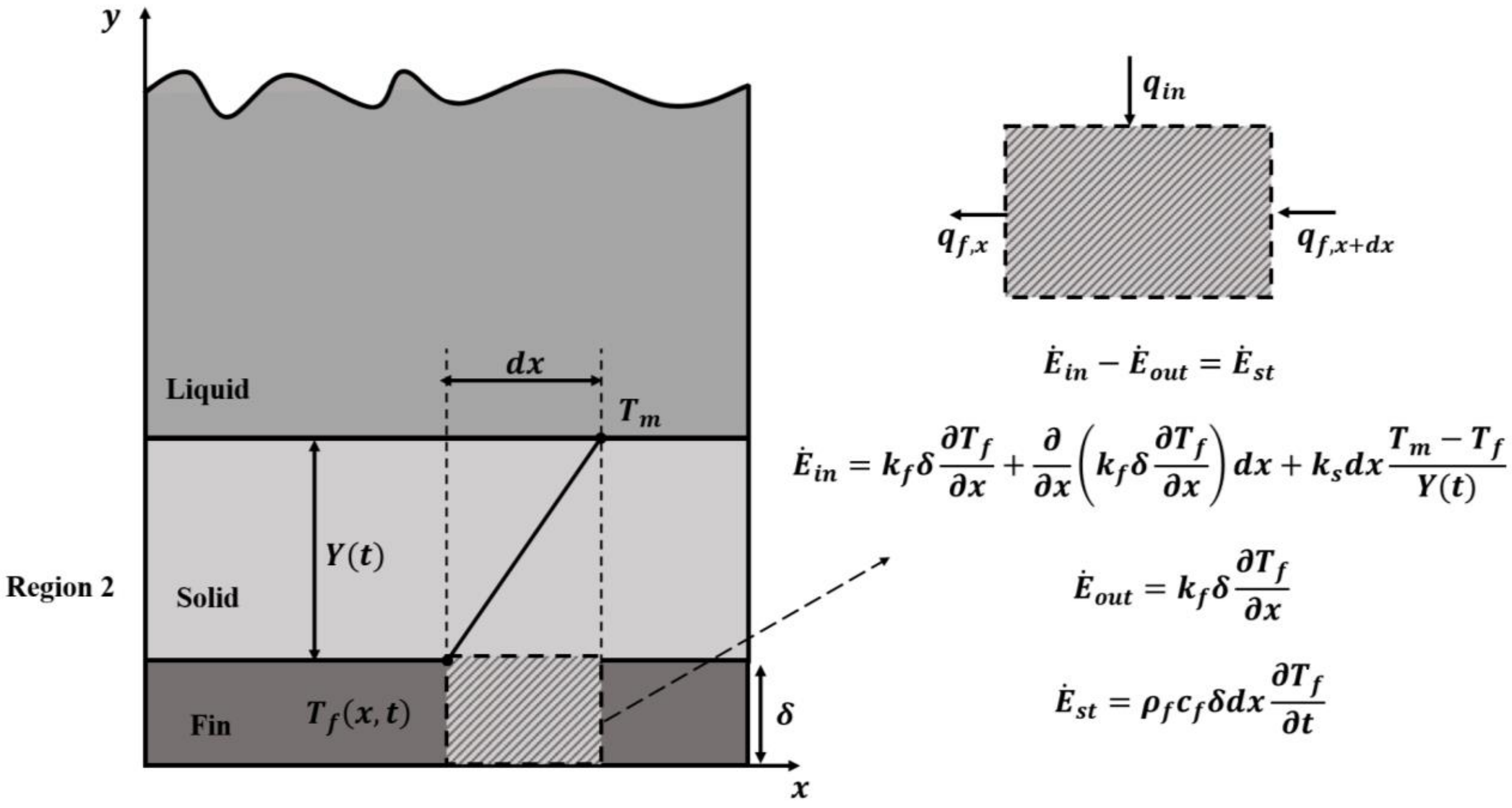
y
q_in
q_f,x
q_f,x+dx
dx
Liquid
T_m
Region 2
Y(t)
Solid
Fin
T_f(x, t)
δ
x
$\dot{E}_{in} - \dot{E}_{out} = \dot{E}_{st}$
$\dot{E}_{in} = k_f \delta \frac{\partial T_f}{\partial x} + \frac{\partial}{\partial x}\left(k_f \delta \frac{\partial T_f}{\partial x}\right) dx + k_s dx \frac{T_m - T_f}{Y(t)}$
$\dot{E}_{out} = k_f \delta \frac{\partial T_f}{\partial x}$
$\dot{E}_{st} = \rho_f c_f \delta dx \frac{\partial T_f}{\partial t}$

(d)

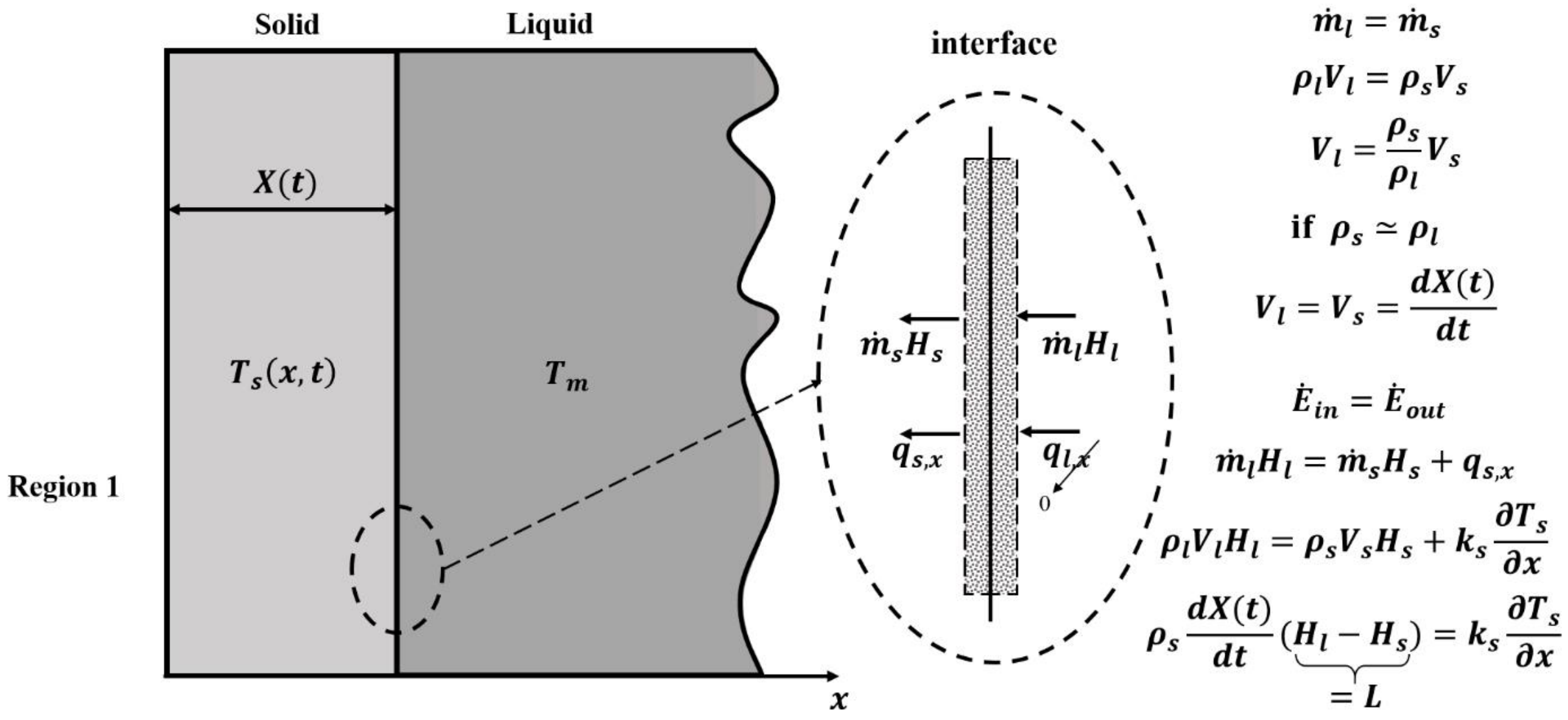

**Figure S1.** (a) Schematic diagram of the one-dimensional problem modeling, depicting the control volume defined within each sub-region. (b) Depiction of the solid, two-phase, and liquid regions in the PCM problem, with the assumption of a temperature change occurring only within the solid region. (c) Derivation of the temperature equation for the fin, based on the assumption of a linear temperature distribution across the thin solid layer formed on the fin. (d) Application of the mass and energy conservation equations to establish the boundary condition for the solid phase undergoing phase change.

Based on the control volume illustrated in Figure S1(a-c), the energy equations for the solid phase and the fin within their respective sub-domains can be simplified as presented in Equations (S.1) and (S.2), respectively.

$$\frac{\partial^2 T_s}{\partial x^2} = \left(\frac{\rho c}{k}\right)_s \frac{\partial T_s}{\partial t} \quad \text{(S.1)}$$

$$\frac{\partial^2 T_f}{\partial x^2} - \frac{k_s}{\delta k_f Y(t)}\left(T_f - T_m\right) = \left(\frac{\rho c}{k}\right)_f \frac{\partial T_f}{\partial t} \quad \text{(S.2)}$$

In addition to the aforementioned equations, the required initial conditions can be specified by Equations (S.2) to (S.6), and the boundary conditions by Equations (S.7) to (S.10). A critical point to note is that in Region 1, the moving boundary is characterized solely by $X(t)$, which is a function of time, whereas in Region 2, it is characterized by $Y(t)$. Utilizing this system of equations along with the specified initial and boundary conditions enables the derivation of the analytical solution presented in the Results section.

$$T_s(x,0) = T_m \tag{S.3}$$

$$T_f(x,0) = T_m \tag{S.4}$$

$$X(0) = 0 \tag{S.5}$$

$$Y(0) = 0 \tag{S.6}$$

To derive the boundary condition for the solid phase undergoing phase change, the conservation equations can be formulated as illustrated in Figure S1 (d). By assuming equal densities for the liquid and solid phases, the equation can be simplified (Equation (S.9)).

$$k_s \frac{\partial T_s(0,t)}{\partial x} = h\left(T_s(0,t) - T_\infty\right) \tag{S.7}$$

$$k_f \frac{\partial T_f(0,t)}{\partial x} = h\left(T_f(0,t) - T_\infty\right) \tag{S.8}$$

$$k_s \frac{\partial T_s(X(t),t)}{\partial x} = \rho_s L \frac{dX(t)}{dt} \tag{S.9}$$

$$-k_f \frac{\partial T_f(l_f,t)}{\partial x} = h\left(T_f(l_f,t) - T_\infty\right) \tag{S.10}$$